%% file: alicepreprint_CDS.tex
\documentclass[ALICE,manyauthors]{cernphprep}

\usepackage[comma,square,numbers,sort&compress]{natbib}
\usepackage{hyperref}
\usepackage{lineno}
\usepackage{esvect}

\newcommand{\vtwo}{\ensuremath{v_{\rm 2}}} 
\newcommand{\pt}{\ensuremath{p_{\rm T}}} 
\newcommand{\vtwohfe}{\ensuremath{v_{\rm 2}^{e^{\pm} \longleftarrow \rm
      HF}}} 
\newcommand{\vtwohfm}{\ensuremath{v_{\rm 2}^{\mu^{\pm} \longleftarrow \rm
      HF}}} 
\newcommand{\vtwob}{\ensuremath{v_{\rm 2}^{\rm Bkg}}} 
\newcommand{\vtwoi}{\ensuremath{v_{2}^{e^{\pm}}}} 
\newcommand{\Rsb}{\ensuremath{R_{\rm SB}}} 
\newcommand{\energy}{\ensuremath{{\sqrt{s_{\mathrm{NN}}}}}} 

\begin{document}%

\begin{titlepage}
\PHyear{2016}
\PHnumber{136}      
\PHdate{26 May}  
%

\title{Elliptic flow of electrons from heavy-flavour hadron decays at
  mid-rapidity in Pb--Pb collisions at $\mathbf{\sqrt{s_\mathrm{NN}}}$ = 2.76 TeV}
\ShortTitle{Elliptic flow of electrons from heavy-flavour hadron decays}   

\Collaboration{ALICE Collaboration\thanks{See Appendix~\ref{app:collab} for the list of collaboration members}}
\ShortAuthor{ALICE Collaboration} 

\begin{abstract}
The elliptic flow of electrons from heavy-flavour hadron decays at mid-rapidity ($|y|$ $<$ 0.7) is
measured in Pb--Pb collisions at \energy~ = 2.76 TeV with  ALICE at
the LHC. The particle azimuthal distribution with respect to the reaction plane can be parametrized
with a Fourier expansion, where the second coefficient ($v_{\rm 2}$) represents the elliptic flow.
The $v_{\rm 2}$ coefficient of inclusive electrons is measured in three centrality classes  (0--10\%,
10--20\% and 20--40\%) with the event plane and the scalar product methods in the transverse momentum (\pt)
intervals 0.5--13 GeV/$c$ and 0.5--8 GeV/$c$, respectively. After
subtracting the background, mainly from photon conversions and Dalitz decays of neutral mesons, a positive $v_{\rm 2}$ of electrons from heavy-flavour hadron
decays is observed in all
centrality classes, with a maximum significance of $5.9\sigma$ in the interval
2 \mbox{$<$ $p_{\rm T}$ $<$ 2.5 GeV/$c$} in semi-central collisions
(20--40\%). The value of $v_{\rm 2}$ decreases towards more
central collisions at low and intermediate $p_{\rm T}$ (0.5 $<$ $p_{\rm T}$ $<$
3 GeV/$c$). The $v_{\rm 2}$ of electrons from heavy-flavour hadron decays at mid-rapidity is found to be
similar to the one of muons from heavy-flavour hadron decays at forward rapidity
\mbox{(2.5 $<$ $y$ $<$ 4)}. The results are described within
uncertainties by model calculations including substantial elastic interactions of heavy quarks with an expanding strongly-interacting medium.

\end{abstract}
\end{titlepage}
\setcounter{page}{2}

\input{Intro.tex}               
\input{ExperimentalApparatus.tex}               
\input{DataAnalysis.tex}               
\input{Results.tex}               
\input{Conclusion.tex}

%
%

\newenvironment{acknowledgement}{\relax}{\relax}
\begin{acknowledgement}
\section*{Acknowledgements}
\input{acknowledgements.tex}    
\end{acknowledgement}

\bibliographystyle{utphys}
\bibliography{biblio}

\newpage
\appendix
\section{The ALICE Collaboration}
\label{app:collab}
\input{Alice_Authorlist_2016-05-05.tex}  
\end{document}

%% file: Intro.tex
\section{Introduction}

The main goal of the ALICE \cite{1748-0221-3-08-S08002} experiment is the study of strongly-interacting matter at the high energy density and temperature reached in ultra-relativistic heavy-ion collisions at the Large Hadron Collider (LHC). In these collisions the formation of a deconfined state of quarks and gluons, the Quark-Gluon Plasma (QGP), is predicted by Quantum ChromoDynamic (QCD) calculations on the lattice~\cite{pQCD1,pQCD2,pQCD3,pQCD4,pQCD5}. Because of their large masses, heavy quarks, i.e. charm ($c$) and beauty ($b$) quarks, are produced at the initial stage of the collision, almost exclusively in hard partonic scattering processes. Therefore, they interact with the medium in all phases of the system evolution, propagating through the hot and dense medium and losing energy via radiative~\cite{Radiativea,Radiativeb} and collisional scattering~\cite{Colla,Collb,Collc} processes. Heavy-flavour hadrons and their decay products are thus effective probes to study the properties of the medium created in heavy-ion collisions. 

Heavy-quark energy loss in strongly-interacting matter can be studied via the modification of the transverse momentum ($p_{\rm T}$) spectra of heavy-flavour hadrons and their decay products in heavy-ion collisions
with respect to the proton-proton yield scaled by the number of binary nucleon-nucleon collisions, quantified by the nuclear modification factor ($R_{\rm{AA}}$).
A strong suppression of open charm hadrons and heavy-flavour decay leptons is observed for $p_{\rm T}$ $>$ 3 GeV/$c$ in central collisions, both at RHIC (\energy~ = 200 GeV)~\cite{STARRaaD0AuAu,STARRaaeAuAu,PHENIXRaaeAuAu,PHENIXRaaeAuAubis,PHENIXRaamuCuCu}  and LHC (\energy ~= 2.76 TeV)~\cite{ALICEDRPbPb,ALICEDRPbPbcent,ALICEmuRPbPb,Adam:2015sza}  energies. The PHENIX and STAR Collaborations measured a $R_{\rm AA}$ of about 0.25 at $p_{\rm T}$ = 5~GeV/$c$ for electrons from heavy-flavour hadron decays at mid-rapidity in central Au--Au collisions at \energy~ = 200 GeV~\cite{STARRaaeAuAu,PHENIXRaaeAuAu,PHENIXRaaeAuAubis}. In addition a similar $R_{\rm AA}$ for D$^{0}$ mesons was measured by STAR~\cite{STARRaaD0AuAu}.
Similar values were measured by the ALICE Collaboration in central Pb--Pb collisions at the LHC for prompt D mesons at mid-rapidity and for muons from heavy-flavour hadron decays at forward rapidity~\cite{ALICEDRPbPb,ALICEDRPbPbcent,ALICEmuRPbPb}. The $p_{\rm T}$ and centrality distributions of the D meson $R_{\rm{AA}}$ are compatible, within uncertainties, with those of charged pions~\cite{ALICEDRPbPbcent}. In addition, the modification of the $p_{\rm T}$ spectra is studied separately for beauty and charm via the $R_{\rm AA}$ of D mesons and non-prompt ${\rm J}/\psi$ from beauty hadron decays measured by the ALICE~\cite{ALICEDRPbPbcent} and CMS Collaborations~\cite{CMSnpjpsi,CMSnpjpsibis}, respectively. A hint for a smaller suppression for beauty than for charm hadrons is observed at high $p_{\rm T}$ in central Pb--Pb collisions, which is well reproduced by calculations including a mass dependence of the parton energy loss~\cite{Dokshitzer:2001zm,Djordjevic:2014,2016PhRvC..93c4904A}.




Further insight into the transport properties of the medium is provided by the measurement of the azimuthal anisotropy of heavy-flavour hadrons and heavy-flavour decay leptons with respect to the reaction plane, defined by the beam axis and the impact parameter of the nucleus--nucleus collision. In non-central  collisions, the initial geometrical anisotropy in coordinate space of the nucleons participating in the collision is converted, by the interactions among the medium constituents, to a final anisotropy in momentum space of the produced particles. This effect can be characterized by the elliptic flow \vtwo, which is the second order harmonic coefficient of the Fourier expansion of the particle azimuthal distribution~\cite{Voloshin:1994mz}. At low $p_{\rm T}$ the measured large $v_{\rm 2}$ of light-flavour hadrons~\cite{lighthadronv23,Abelev:2014pua,lighthadronv25,lighthadronv26} is considered as an evidence for the collective hydrodynamical expansion of the medium~\cite{Evidencehydro,Evidencehydrobis}. 
On general theoretical ground, the formation time of heavy quarks, shorter than 1/(2\,$m_{\rm c, b}$) where $m$ is the mass of the quark ($\approx$ 0.08 fm/$c$ for charm), is expected to be smaller than the QGP thermalization time ($\approx$0.6--1\,fm/$c$~\cite{Liu:2012ax}) with a very small annihilation rate~\cite{Annihalition}.
The heavy-flavour elliptic flow measurements carry information about their degree of thermalization and participation to the collective expansion of the system. It is also relevant for the interpretation of recent results on ${\rm J}/\psi$ anisotropy~\cite{ALICE:Jpsi}, because the ${\rm J}/\psi$ mesons formed from charm quarks in a deconfined partonic phase are expected to inherit the azimuthal anisotropy of their constituent quarks~\cite{Model:Jpsi1,Model:Jpsi2}. At low and intermediate $p_{\rm T}$, the $v_{2}$ of heavy-flavour hadrons and their decay products is also expected to be sensitive to the heavy-quark hadronisation mechanism. Hadronisation via the recombination of heavy quarks with light quarks from the thermalized medium could further increase the elliptic flow of heavy-flavour hadrons and their decay products~\cite{Recombinationa,Recombinationb,Batsouli:2002qf}. At high $p_{\rm T}$  the $v_{2}$ measurements can constrain the path-length dependence of the in-medium parton energy loss, which is different for radiative~\cite{Radiativea,Radiativeb} and collisional~\cite{Colla,Collb,Collc} energy loss mechanisms. Particles emitted in the direction of the reaction plane have, on average, a shorter in-medium path length than those emitted orthogonally to it, leading to an expected positive elliptic flow~\cite{Pathlengtha,Pathlengthb}, as observed for charged hadrons~\cite{lighthadronv21,lighthadronv22,lighthadronv23,ALICEChpartv2,lighthadronv25,lighthadronv26}.



At RHIC, a positive elliptic flow of heavy-flavour decay electrons at low and intermediate $p_{\rm T}$ was reported by the PHENIX and STAR Collaborations~\cite{PHENIXRaaeAuAu,Adamczyk:2014yew}  at mid-rapidity in Au--Au collisions at  $\sqrt{s_\mathrm{NN}}$ = 200 GeV, reaching a maximum value of about 0.15 at $p_{\rm T}$ = 1.5 GeV/$c$ in semi-central collisions. Elliptic flow values measured at lower colliding energies are found to be consistent with zero~\cite{Adamczyk:2014yew}. The ALICE Collaboration measured the elliptic flow of D mesons at mid-rapidity~\cite{Abelev:2013lca,Abelev:2014ipa} and heavy-flavour decay muons at forward rapidity~\cite{Adam:2015pga} in Pb--Pb collisions at $\sqrt{s_\mathrm{NN}}$ = 2.76 TeV. At intermediate $p_{\rm T}$ a positive $v_{\rm 2}$ of prompt D mesons (5.7$\sigma$ effect in the interval 2 $<$ $p_{\rm T}$ $<$ 6 GeV/$c$ for the 30--50\% centrality class), and heavy-flavour decay muons (3$\sigma$ effect in the interval 3 $<$ $p_{\rm T}$ $<$ 5 GeV/$c$ for the 10--20\% and 20--40\% centrality classes) is observed. The centrality dependence shows a hint for a decrease of $v_{\rm 2}$ towards central collisions. At high $p_{\rm T}$ ($p_{\rm T}$ $>$ 8\, GeV/$c$ for D mesons and $p_{\rm T}$ $>$ 6 GeV/$c$ for heavy-flavour decay muons) small values of $v_{\rm 2}$ are measured, compatible with zero within large uncertainties. 

We report on the measurement of the elliptic flow of electrons from heavy-flavour hadron decays at mid-rapidity ($|$$y$$|$ $<$ 0.7) in Pb--Pb collisions at $\sqrt{s_\mathrm{NN}}$ = 2.76 TeV with ALICE. The measurement is performed in the $p_{\rm T}$ interval 0.5 $<$ $p_{\rm T}$ $<$ 13 GeV/$c$ in three centrality classes 0--10\%, 10--20\% and 20--40\% with the event plane method.  
The results complement the heavy-flavour decay muon $v_{2}$ measurements at forward rapidity~\cite{Adam:2015pga} and extend towards lower $p_{\rm T}$ those of D mesons at mid-rapidity~\cite{Abelev:2013lca}. Moreover, charm hadron decays are expected to mainly contribute to the heavy-flavour decay electron sample at low $p_{\rm T}$ ($p_{\rm T}$ $<$ 3\,GeV/$c$), whereas at higher $p_{\rm T}$ the contribution from beauty hadron decays should become relevant~\cite{Abelev:2014hla,Abelev:2012sca}. Therefore, the measurement of heavy-flavour decay electron $v_{\rm 2}$ provides further inputs on the beauty and charm elliptic flow at mid-rapidity to theoretical calculations that aim at describing the heavy-quark interactions with the medium. 
The elliptic flow of inclusive electrons obtained with the scalar product method is also compared to the measurements performed with the event plane method to study possible non-flow contributions and biases due to the method itself.


This article is organized as follows: the experimental apparatus and data sample used in the analysis are presented in Section~\ref{section2}. The analysis strategy, including the electron identification and the procedure for the subtraction of the background due to electrons not originating from heavy-flavour hadron decays, are described in Section~\ref{section3}. The elliptic flow of heavy-flavour decay electrons is presented in Section~\ref{results} and compared to theoretical models in Section~\ref{modelcomparison}. The  summary and conclusions of this article are presented in Section~\ref{section6}.


%% file: ExperimentalApparatus.tex
\section{Experimental apparatus and data sample} \label{section2}

The ALICE experimental apparatus is described in detail in~\cite{1748-0221-3-08-S08002,Abelev:2014ffa}. 
The global reference system has the $z$-axis parallel to the beam line, the $x$-axis pointing towards the centre of the LHC accelerator ring and the $y$-axis pointing upward. 
In the following, the subsystems that are relevant for the heavy-flavour decay electron analysis are described.

Charged particle tracks are reconstructed at mid-rapidity ($|$$\eta$$|$ $<$ 0.9) in the central barrel of ALICE with the Time Projection Chamber (TPC) and the Inner Tracking System (ITS). 
The electron identification uses information from the ITS, TPC and the Time-of-Flight (TOF) detectors in the $p_{\rm T}$ interval 0.5 $<$ $p_{\rm T}$ $<$ 3 GeV/$c$ and from the TPC and ElectroMagnetic Calorimeter (EMCal) in the $p_{\rm T}$ interval 3 $<$ $p_{\rm T}$ $<$ 13 GeV/$c$. In the following, the two identification methods will be referred to as ITS-TPC-TOF and TPC-EMCal analyses, respectively. 
These detectors are located inside a large solenoidal magnet that provides a uniform magnetic field of 0.5 T along the beam direction. The event characterization is performed with two scintillator detectors, V0, used for triggering, centrality and reaction plane estimation. Together with the Zero Degree Calorimeters (ZDC), they are used to further select events offline.

The ITS~\cite{1748-0221-5-03-P03003} detector consists of six cylindrical silicon layers surrounding the beam vacuum tube. 
The first two layers are positioned at 3.9 and 7.6\,cm radial distance from the beam line. 
Dealing with the high particle density in this region requires an excellent position resolution, which is achieved with Silicon Pixel Detectors (SPD). 
The third and fourth layers are radially positioned at 15 and 23.9\,cm and consist of Silicon Drift Detectors (SDD), while the two outermost layers are radially positioned at 38 and 43 cm and are made of Silicon Strip Detectors (SSD). The four SDD and SSD layers enable charged-particle identification via the measurement of their energy loss d$E$/d$x$ with a resolution of about 10--15\%. 

The TPC~\cite{Alme:2010ke} detector has a cylindrical shape with an inner radius of about 85 cm, an outer radius of about 250 cm, and a length of 500 cm. 
The TPC is the main tracking detector of the central barrel and is optimized to provide, together with the other central barrel detectors, charged-particle momentum measurement with excellent two-track separation and particle identification. For a particle traversing the TPC, up to 159 space points are recorded and used to estimate its specific energy loss. The resolution of the  d$E$/d$x$ measured in the TPC is approximately 6\% for minimum-ionizing particles passing through the full detector.

At a radial distance of 3.7\,m from the beam axis, the TOF detector~\cite{tofperf} improves further the particle identification capability of ALICE. It provides a measurement of the time of flight for the particles from the interaction point up to the detector itself with an overall resolution of about 80\,ps for pions and kaons at $p_{\rm T}$ $=$ 1\,GeV/$c$ in the Pb--Pb collision centrality intervals used in this analysis. 
The measured time-of-flight of electrons is well separated from those of kaons and protons up to $p_{\rm T}$ $\simeq$ 2.5 GeV/$c$ and $p_{\rm T}$ $\simeq$ 4 GeV/$c$, respectively.

The EMCal~\cite{2010arXiv1008.0413R} is a Pb-scintillator sampling calorimeter located at a radial distance of about 4.5\,m from the beam axis spanning the pseudorapidity range $|$$\eta$$|$$<$ 0.7 and covering 107$^{\circ}$ in azimuth. The cell size of the EMCal is approximately 0.014 rad $\times$0.014 in $\Delta \varphi$ $\times$ $\Delta \eta$. The energy resolution has been measured to be 1.7$\oplus$11.1/$\sqrt{E\rm(GeV)}$$\oplus$5.1/$E$(GeV)\%. The EMCal increases the existing ALICE capabilities to measure high-momentum electrons.

The V0 detectors~\cite{vZero} consist of two arrays of 32 scintillator tiles covering the pseudorapidity ranges 2.8 $<$ $\eta$  $<$ 5.1 (V0A) and \mbox{$-$3.7 $<$ $\eta$ $<$ $-$1.7} (V0C), respectively. 
The two arrays are arranged in four rings each around the beam pipe. 
The V0 detectors are used to select beam--beam interactions online. For Pb--Pb collisions, the total signal amplitude is fitted with a model based on the Glauber approach, which is used to classify events according to their centrality classes~\cite{centralitypaper}, which correspond to percentiles of the hadronic cross section. For instance, the 0--10\% centrality class corresponds to the 10\% most central events. In addition, the azimuthal segmentation of the V0 detectors allows for an estimation of the reaction plane direction. 

The ZDCs~\cite{Arnaldi:409740} are located on both sides of the interaction point at z $\approx$ $\pm$114\,m. Parasitic collisions of main bunches with satellite bunches are rejected on the basis of the timing information from the neutron ZDCs. 

The results presented in this paper are based on a data sample of Pb--Pb collisions recorded with ALICE in November and December 2011 at \energy~ = 2.76\,TeV. The events were collected with a minimum-bias interaction trigger using information of the coincidence of signals between V0A and V0C detectors.
Central and semi-central Pb--Pb collisions were selected online by applying thresholds on the V0 signal amplitudes resulting in two separate trigger classes (central and semi-central triggers). In addition to the central and semi-central data samples, events selected by the EMCal trigger are analysed. The EMCal trigger required an EMCal cluster energy summed over a group of 4$\times$4 cells, implemented as a sliding window, larger than an energy threshold. A centrality-dependent energy threshold was used, varying approximately from 7 GeV in the 0--10\% centrality class to 2 GeV in the \mbox{80--90\%} centrality class. The EMCal trigger is in coincidence with the minimum-bias trigger. Detailed trigger information for the ALICE apparatus are reported in~\cite{Abelev:2014ffa}. 

Only events with a reconstructed interaction vertex (primary vertex), determined by extrapolating charged-particle tracks to the beam line, with $|z|$ $<$ 10 cm from the nominal interaction point  are used in the analysis in order to minimize edge effects at the limit of the central barrel acceptance. 
In addition, the $z$ position of the primary vertex reconstructed using tracklets defined by hit pairs in the SPD is required to agree within 0.5\,cm with the one of the primary vertex reconstructed with tracks.
Since the \vtwo~measurements could be biased by multiplicity outliers, the centrality estimated with the V0 information is compared to that estimated using the number of reconstructed tracks in the TPC. 
Events with an absolute difference between the centrality estimated with the V0 detectors and the one estimated with the TPC detector larger than 5\%, corresponding to events with pile-up from different bunch crossings, are rejected from the analysis. 
The event selection removed about 5\% of the total number of events depending on the trigger and the centrality of Pb--Pb collisions. The number of events analysed after applying the event selection are listed in Table~\ref{table:event_cutddd} for the different centrality classes and triggers together with the corresponding integrated luminosity.  The EMCal trigger is not used in the 0--10\% centrality class because of the high statistics achieved with the central trigger.

\begin{table}
\begin{center}
\centering
\begin{tabular}{cccc}
\hline 
Centrality class & Trigger system & $N_{\rm events}$ & $L_{\rm int}$ ($\mu$b$^{-1}$)\\ 
\hline
0--10\%  & Central trigger &15$\times$$\mathrm{10^6}$ & 19.6 \\
10--20\%  & Semi-central trigger  & 4$\times$$\mathrm{10^6}$ &  5.2 \\
 20--40\%  & Semi-central trigger  & 8$\times$$\mathrm{10^6}$ &  5.2 \\
 
 \hline
 10--20\%  & EMCal trigger & 0.7$\times$ $\mathrm{10^6}$ &  29.1\\
 20--40\%  & EMCal trigger & 1$\times$ $\mathrm{10^6}$ &   24.4\\
\hline 
\end{tabular}
\caption[Event selection]{Number of events and integrated luminosity for the different triggers (see text) and centrality classes considered in this analysis. The centrality classes are expressed as percentiles of the hadronic cross section~\cite{centralitypaper}.}
\label{table:event_cutddd}
\end{center}
\end{table}

%% file: DataAnalysis.tex
\section{Data analysis} \label{section3}

The elliptic flow of electrons from heavy-flavour hadron decays \vtwohfe~ is obtained from the measurement of the inclusive electron elliptic flow \vtwoi~by subtracting the elliptic flow of electrons which do not originate from heavy-flavour hadron decays, \vtwob.
Exploiting the additive property of the particle azimuthal angle distribution with respect to the reaction plane, \vtwohfe~ can be expressed as:

\begin{equation}
\vtwohfe = \dfrac{(1+\Rsb) \vtwoi - \vtwob}{\Rsb},
\label{hfeflow}
\end{equation}

where $R_{\rm SB}$ is the ratio of the heavy-flavour decay electron yield to that of background electrons. In this paper, electrons from heavy-flavour hadron decays include electrons from quarkonium decays, whose contribution is however expected to be small as discussed in Section~\ref{NHFE}. In the following sections, the \vtwoi and  $R_{\rm SB}$ measurements are presented, as well as the two procedures to determine \vtwob. 

\subsection{Track selection and electron identification}

\vspace{0.5 cm}

 \begin{table}[ht!]
\begin{center}
\centering
   \begin{tabular}{ccc}
\hline 
Analysis  & ITS-TPC-TOF & TPC-EMCal \\
\hline
$p_{\rm T}$ range (GeV/$c$) & 0.5--3 & 3--13 \\
$|y|$ & $<$ 0.8 & $<$ 0.7 \\
\hline
Number of TPC points &  $\ge$ 100 & $\ge$ 100 \\
Number of TPC points in d$E$/d$x$ calculation &  $\ge$ 90 &  -- \\
Ratio of found TPC points over findable &  $>$ 0.6 & $>$ 0.6\\
$\chi^2$/point of the momentum fit in the TPC & $<$  3.5 & $<$  3.5 \\
DCA$_{xy}$ & $<$ 2.4 cm & $<$ 2.4 cm\\
DCA$_{z}$& $<$ 3.2 cm & $<$ 3.2 cm\\
Number of ITS hits & $\ge$ 5  & $\ge$ 3  \\
Number of hits in the SPD layers & 2 & $\ge$ 1  \\
\hline 
\end{tabular}
\caption{Summary of the track selection criteria used in the analyses.}
\label{table:track_cuts}
\end{center}
\end{table}

Electron candidate tracks are required to fulfill the track selection criteria summarized in Table~\ref{table:track_cuts}. Tracks are selected by requiring at least 100 associated space points in the TPC with at least 90 used for the d$E$/d$x$ calculation and a value of the $\chi^{2}$/point of the momentum fit in the TPC smaller than 3.5. 
These selection criteria suppress the contribution from short tracks, which are unlikely to originate from the primary vertex. To further reduce the contamination from particles originating either from weak decays of light hadrons or from the interaction of other particles with the detector  material, only tracks with a maximum value of the distance of closest approach (DCA) to the primary vertex in both the $xy$-plane \mbox{(DCA$_{xy}$ $<$ 2.4 cm)} and the $z$ direction \mbox{(DCA$_{z}$ $<$ 3.2 cm)} are accepted. In addition, in order to minimize the contribution of electrons coming from $\gamma$ conversions in the detector material at large radii, hits in both SPD layers are required for all selected tracks in the ITS-TPC-TOF analysis ($p_{\rm T}$ $<$ 3 GeV/$c$). 
Tracks are required to have at least three out of the four possible hits in the external layers of the ITS (SDD and SSD) in order to have at least three d$E$/d$x$ measurements to be used for the Particle IDentification (PID).
This guarantees a good particle identification based on the d$E$/d$x$ in the ITS. 
Since the azimuthal coverage of the EMCal had a significant superposition with parts of the SPD detector that were not active during the data taking, this approach has to be modified for the TPC-EMCal analysis \mbox{($p_{\rm T}$ $>$ 3 GeV/$c$)}. In this case, at least one hit in any of the two SPD layers is required and the minimum number of associated ITS hits is reduced to 3. This results in a larger contribution of conversion electrons in the inclusive electron sample. The signal-to-background ratio is, as a consequence, smaller in the TPC-EMCal analysis than in the ITS-TOF-TPC analysis at the same $p_{\rm T}$. 

\begin{figure}[!ht]
 \centering
\includegraphics[scale=0.37]{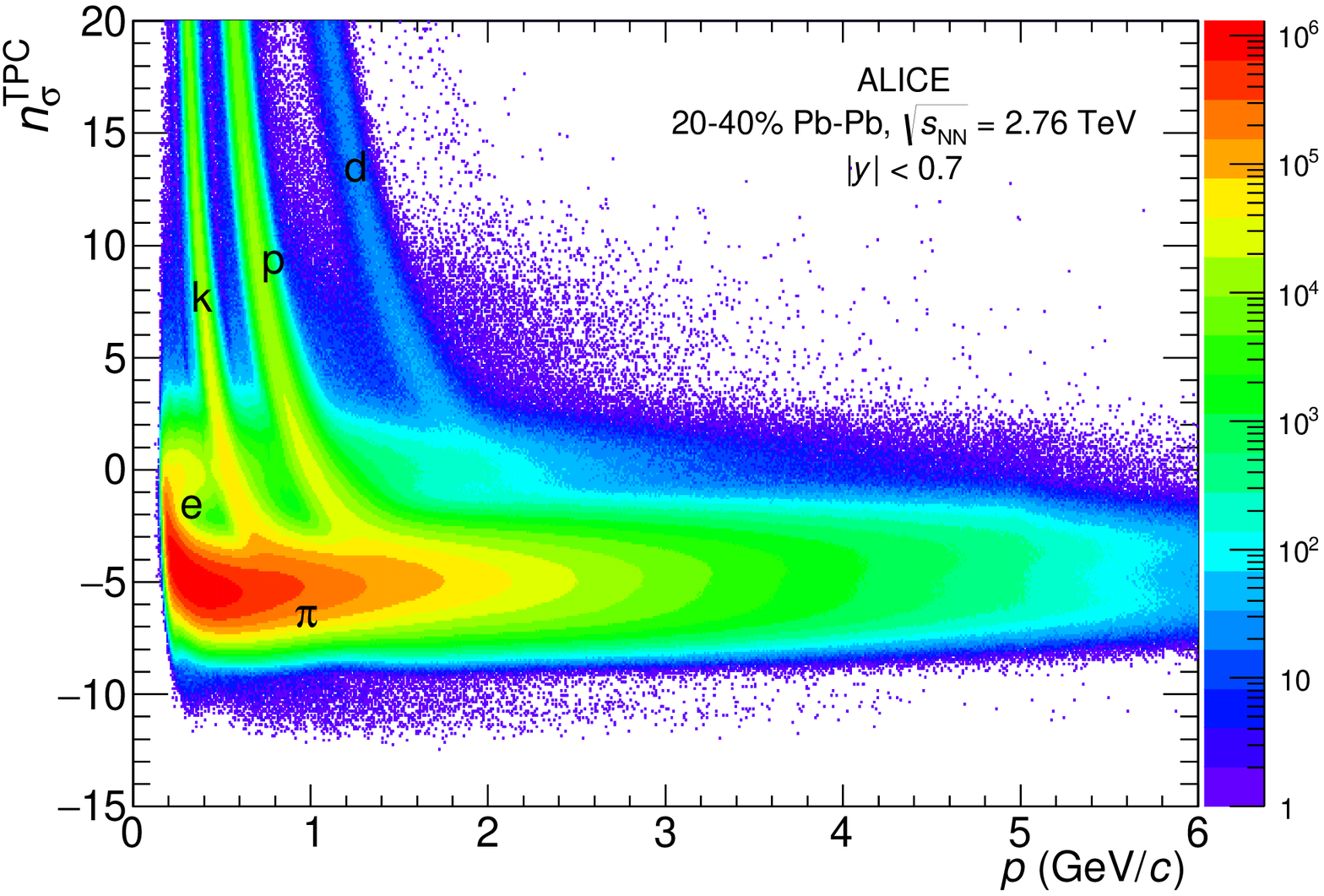}
\includegraphics[scale=0.37]{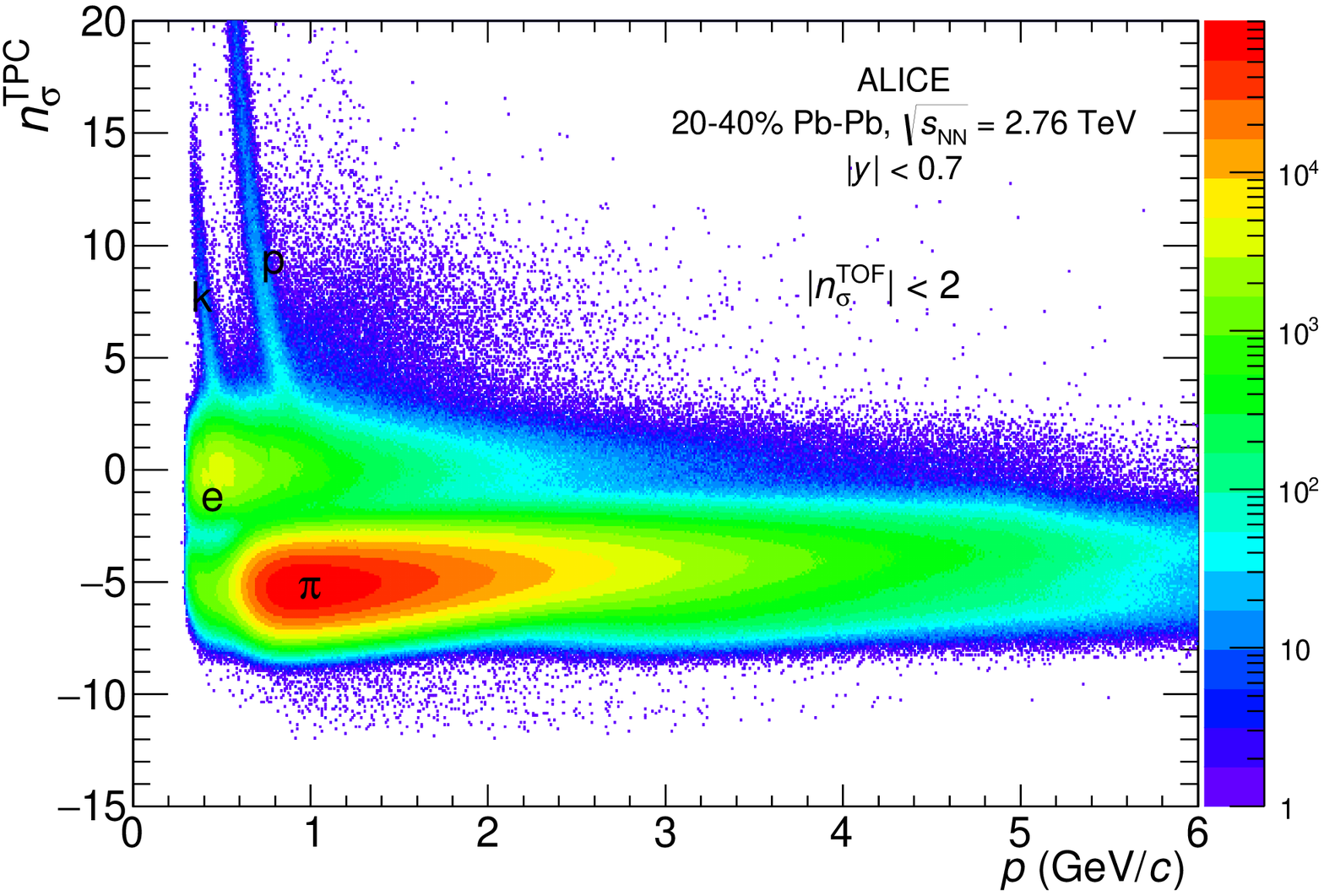}
\vspace{0.5 cm}
\includegraphics[scale=0.37]{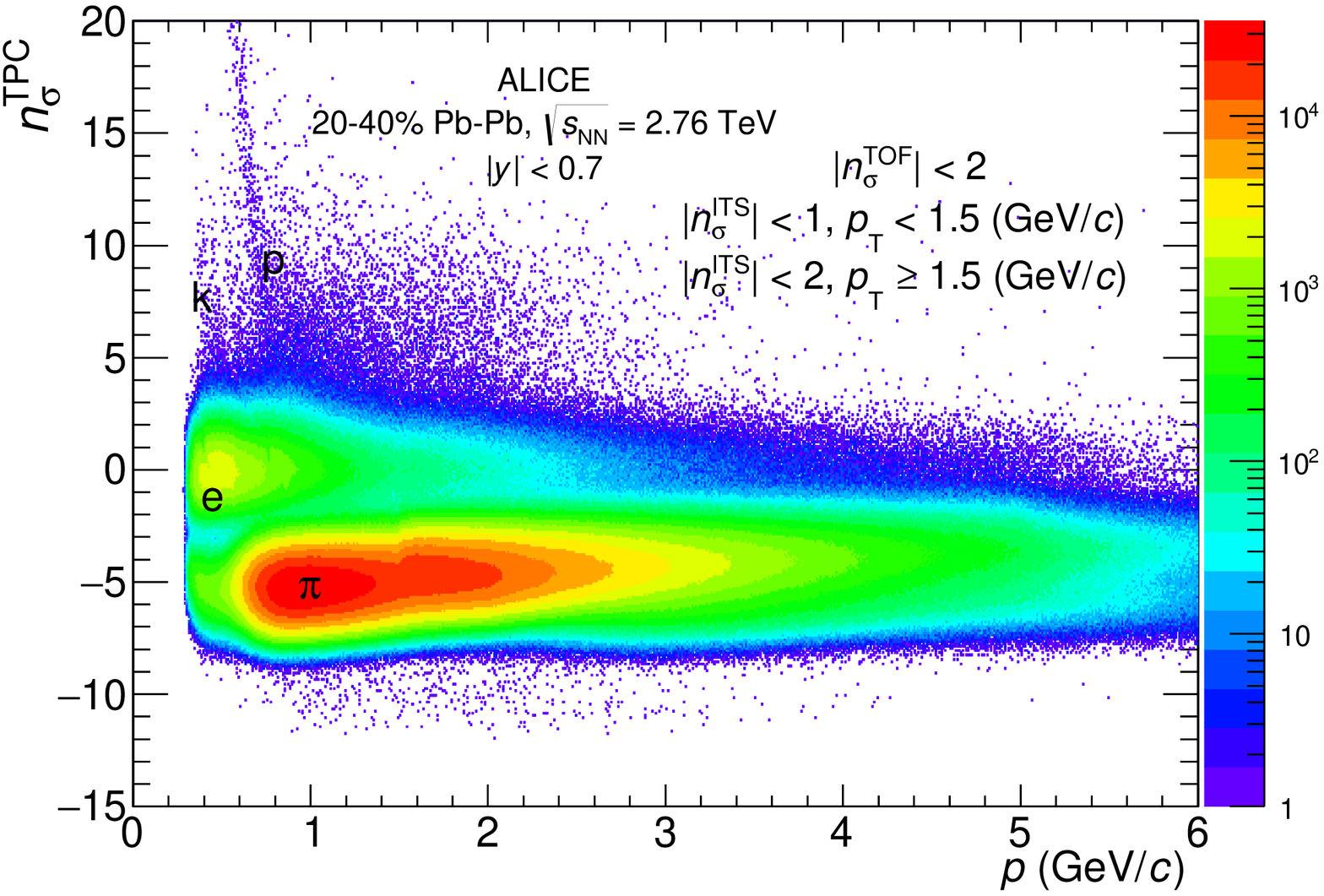}
 \caption{$n_{\sigma}^{\rm{TPC}}$ distributions as a function of momentum in semi-central (20--40\%) Pb--Pb collisions at \energy~ = 2.76 TeV. Upper left panel: no ITS or TOF electron identification is applied. Upper right panel: the TOF-PID (see text) is applied. Lower panel: the TOF and ITS-PID (see text) are both applied.}
 \label{ITSTOFTPCPI}
\end{figure}

\begin{table}[ht!]
\begin{center}
\centering
   \begin{tabular}{ccccc}
\hline 
$p_{\rm T}$ range & TPC d$E$/d$x$ cut  & ITS d$E$/d$x$ cut  & TOF compatibility  & $E$ /$p$ matching \\
(GeV/$c$) &  &  & with $e$ hypothesis  & \\
\hline
\,\, 0.5--1.5 & $-$1 $<$ $n_{\sigma}^{\rm{TPC}}$ $<$ 3 & $|$$n_{\sigma}^{\rm{ITS}}$$|$ $<$ 1 & $|$$n_{\sigma}^{\rm{TOF}}$$|$ $<$ 2 & \\
1.5--3 & \, \, 0 $<$ $n_{\sigma}^{\rm{TPC}}$ $<$ 3 & $|$$n_{\sigma}^{\rm{ITS}}$$|$ $<$ 2 & $|$$n_{\sigma}^{\rm{TOF}}$$|$ $<$ 2 & \\
3--8 & $-$1 $<$ $n_{\sigma}^{\rm{TPC}}$ $<$ 3 &  & & 0.8 $<$ $E$/$p$ $<$ 1.2 \\
8--13 & $-$1 $<$ $n_{\sigma}^{\rm{TPC}}$ $<$ 3 &  & & $-$2 $<$ $n_{\sigma}^{\rm{EMCal}}$ $<$ 3  \\
\hline 
\end{tabular}
\caption{Summary of the electron identification criteria used in the analyses (see text for more details).}
\label{table:pid_cuts}
\end{center}
\end{table}

Electron identification is mainly based on the measurement of the specific energy loss in the TPC (d$E$/d$x$). The discriminant variable used, $n_{\sigma}^{\rm{TPC}}$, is the deviation of this quantity from the parameterized electron Bethe-Bloch~\cite{BetheBloch} expectation value, expressed in units of the d$E$/d$x$ resolution~\cite{Abelev:2014ffa}. This distribution is shown as a function of the track momentum in semi-central triggered events for the 20--40\% centrality class in the upper left panel of Figure~\ref{ITSTOFTPCPI}. In the low momentum region the kaon, proton and deuteron d$E$/d$x$ bands cross that of electrons. In addition, the particle identification at high momentum is limited by the merging of the d$E$/d$x$ bands of electrons, pions, muons and other hadrons, therefore the information of other detectors is mandatory to select a pure sample of electrons. Table~\ref{table:pid_cuts} summarizes the PID cuts.

\begin{figure}[h]
\centering
\includegraphics[scale=0.37]{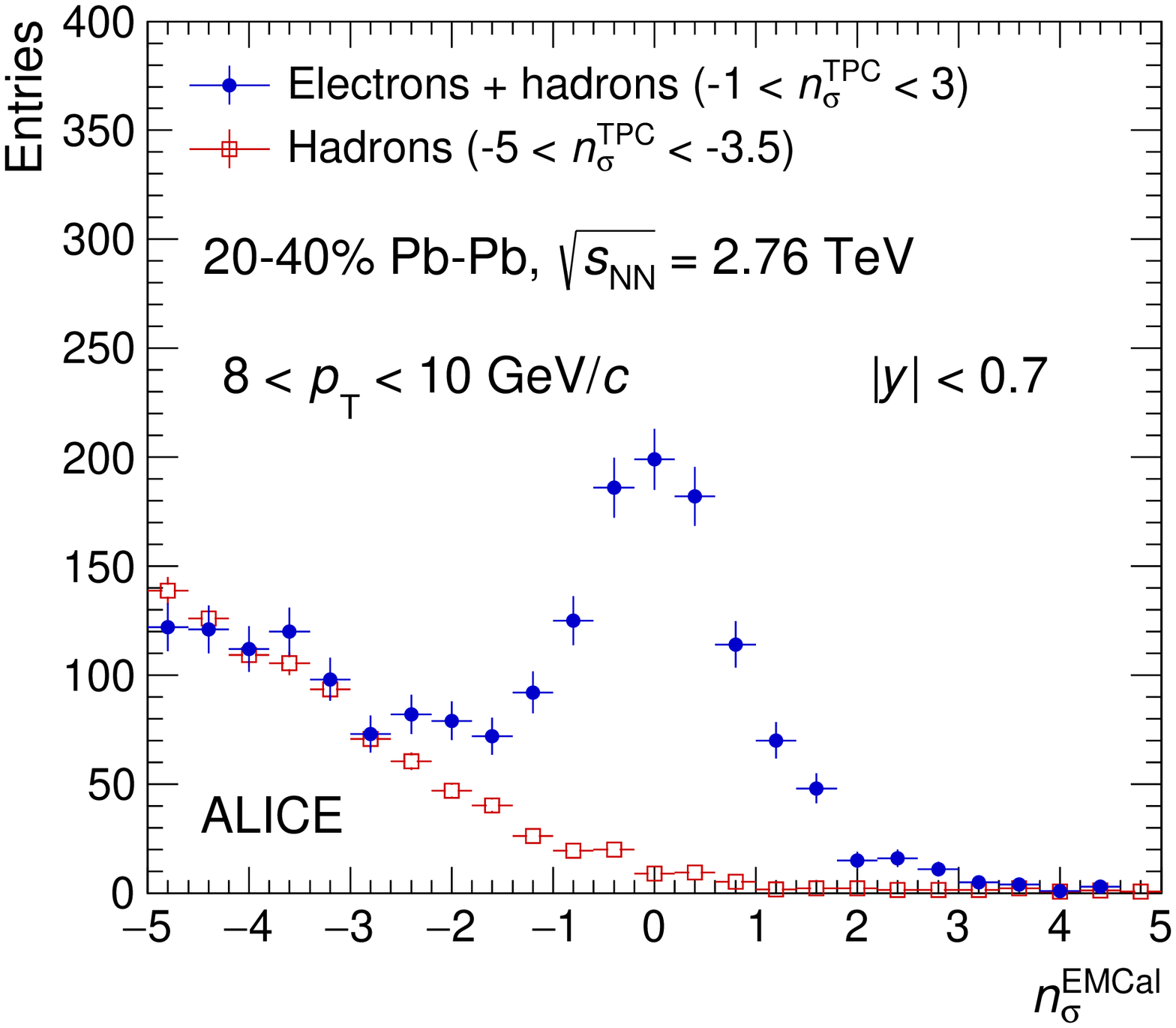}
\caption{Deviation of the measured $E$/$p$ from the expected $\langle$$E$/$p$$\rangle$ of electrons divided by the $E$/$p$ resolution ($n_{\sigma}^{\rm{EMCal}}$) for tracks in the $p_{\rm T}$ interval 8--10 GeV/$c$ in semi-central (20--40\% centrality class) Pb--Pb collisions at \energy~ = 2.76 TeV.  Electron and hadron candidates are selected with the TPC d$E$/d$x$ by requiring $-$1 $<$ $n_{\sigma}^{\rm{TPC}}$ $<$ 3  and $-$5 $<$ $n_{\sigma}^{\rm{TPC}}$ $<$ $-$3.5, respectively.}
\label{eop}
\end{figure}

At low $p_{\rm T}$ (0.5 $<$ $p_{\rm T}$ $<$ 3 GeV/$c$), the measured time-of-flight in the TOF detector and the d$E$/d$x$ in the ITS are used in addition to the TPC d$E$/d$x$ to further reject hadrons. In the top right panel of Figure~\ref{ITSTOFTPCPI}, the $n_{\sigma}^{\rm{TPC}}$ distribution is shown after requiring that the measured time-of-flight of the particle is compatible with the electron hypothesis within two sigmas, where sigma is the time-of-flight resolution ($|$$n_{\sigma}^{\rm{TOF}}$$|$ $<$ 2). The kaon and proton contributions in the low momentum region are reduced but not completely removed due to wrongly associated hits in the TOF detector. 
This source of contamination is further suppressed using the  d$E$/d$x$ in the ITS. 
This selection is applied using the $n_{\sigma}^{\rm{ITS}}$ variable, defined in the same way as for the TPC. Electron candidates are selected with 
$|$$n_{\sigma}^{\rm{ITS}}$$|$ $<$ 1 for 0.5 $<$ $p_{\rm T}$ $<$1.5 GeV/$c$  and with \mbox{$|$$n_{\sigma}^{\rm{ITS}}$$|$ $<$ 2} for 1.5 $<$ $p_{\rm T}$ $<$ 3 GeV/$c$, where the particles species are less separated in $n_{\sigma}^{\rm{ITS}}$.
 In the lower panel of Figure~\ref{ITSTOFTPCPI}, the $n_{\sigma}^{\rm{TPC}}$ distribution is shown after the additional electron identification criteria in the ITS are applied. A pure electron sample is obtained by selecting tracks with $-$ 1 $<n_{\sigma}^{\rm{TPC}}<$ 3 and 0 $<n_{\sigma}^{\rm{TPC}}<$ 3 in the intervals \mbox{0.5 $<$ $p_{\rm T}$ $<$ 1.5 GeV/$c$} and \mbox{1.5 $<$ $p_{\rm T}$ $<$ 3 GeV/$c$}, respectively. In order to keep the contamination below 5\%, the stronger requirement in the $p_{\rm T}$ interval \mbox{1.5 $<$ $p_{\rm T}$ $<$ 3 GeV/$c$} is applied due to the merging of the pion and electron d$E$/d$x$ bands in the TPC.

In the $p_{\rm T}$ interval 3--13 GeV/$c$, the electron identification is based on the measurement of the TPC d$E$/d$x$ and the $E$/$p$ ratio, where $E$ is the energy of the EMCal cluster matched to the prolongation of the track with momentum $p$ reconstructed with the TPC and ITS detectors. Unlike for hadrons, the ratio $E$/$p$ is around 1 for electrons, because they deposit most of their energy in the EMCal. 
In addition, the EMCal cluster shape is used to improve the purity of the electron sample, because the profile of the shower produced by electrons is more circular than the one produced by hadrons \cite{Berger:1992}. 
In the  $p_{\rm T}$ interval 8--13 GeV/$c$,  the EMCal PID selection is applied in terms of $n_{\sigma}^{\rm{EMCal}}$, which is defined as the deviation of the measured $E$/$p$ from the expected $\langle~E$/$p$$\rangle$ for electrons obtained from data and  normalized by the width of the electron $E$/$p$ distribution obtained with a fit Gaussian function. Electron candidates are selected with the identification criteria $-$1 $<$ $n_{\sigma}^{\rm{TPC}}$ $<$ 3 and $-$2 $<$ $n_{\sigma}^{\rm{EMCal}}$ $<$ 3 in the $p_{\rm T}$ interval \mbox{8 $<$ $p_{\rm T}$ $<$ 13 GeV/$c$}.


The hadron contamination in the $p_{\rm T}$ interval 0.5--3 GeV/$c$ is estimated by fitting in momentum slices the TPC d$E$/d$x$ distribution after the TOF- and ITS-PID selections with a convolution of Landau and exponential functions, similarly to what was done in~\cite{Abelev:2012xe}. For $p_{\rm T}$ $>$ 3 GeV/$c$, the hadron contamination is obtained from the $E$/$p$ distribution of reconstructed tracks in momentum slices after applying only the TPC-PID selection. The estimated hadron contamination is lower than 5\% up to $p_{\rm T}$ = 8 GeV/$c$ with negligible dependence on centrality, event plane and pseudorapidity and therefore it is not subtracted. 
The possible effect induced by this contamination is considered in the systematic uncertainties, as discussed in Section~\ref{SystUncInclElecV2}.
For higher $p_{\rm T}$ (\mbox{8 $<$ $p_{\rm T}$ $<$ 13 GeV/$c$}), the contamination of hadrons is subtracted statistically from the electron sample in the $n_{\sigma}^{\rm{EMCal}}$  distributions before \mbox{calculating \vtwoi.}  
The $n_{\sigma}^{\rm{EMCal}}$ distribution for tracks in the $p_{\rm T}$ interval 8 $<$ $p_{\rm T}$ $<$ 10 GeV/$c$ in semi-central (20--40\%) Pb--Pb collisions at \energy~ = 2.76 TeV is shown in Figure~\ref{eop}. Electrons and hadrons candidates are selected with the TPC d$E$/d$x$ by requiring \mbox{$-$1 $<$ $n_{\sigma}^{\rm{TPC}}$ $<$ 3} and \mbox{$-$5 $<$ $n_{\sigma}^{\rm{TPC}}$ $<$ $-$3.5}, respectively. 
The $n_{\sigma}^{\rm{EMCal}}$ distribution of hadrons is scaled to the $n_{\sigma}^{\rm{EMCal}}$ distribution of electron candidates in the range \mbox{$-$5 $< n_{\sigma}^{\rm{EMCal}} <$ $-$3} to determine statistically the amount of hadrons after the TPC-PID selection.  The subtracted contamination of hadrons reaches approximately 15\% and 20\% in the $p_{\rm T}$ intervals 8  $<$ $p_{\rm T}$ $<$ 10 GeV/$c$ and 10  $<$ $p_{\rm T}$ $<$ 13 GeV/$c$, respectively, in all centrality classes.

The rapidity ranges used in the ITS-TPC-TOF ($p_{\rm T}$ $<$ 3 GeV/$c$) and TPC-EMCal ($p_{\rm T}$ $>$ 3 GeV/$c$) analyses are restricted to $|y|$ $<$ 0.8 and $|y|$ $<$ 0.7, respectively, to avoid the edges of the detectors, where the systematic uncertainties related to particle identification increase. It was checked, by restricting the ITS-TPC-TOF analysis to $|y|$ $<$ 0.7, that the change in the results due to the different $y$ range are not significant. In the following the notation $|y|$ $<$ 0.7 will be used.

\subsection{Flow methods}\label{FlowMethods}

The $p_{\rm T}$-differential azimuthal distribution of produced particles can be described by a Fourier expansion of the Lorentz invariant distribution of outgoing momenta~\cite{Voloshin:1994mz}:

\begin{equation}
E \frac{{\rm d}^3 N}{{\rm d}p^{3}} = \frac{1}{2 \pi} \frac{{\rm d}^2 N}{\pt {\rm d}\pt {\rm d}y} \Bigg( 1 + \sum^{\infty}_{n=1} 2 v_{n} \cos [n(\varphi - \Psi_{n})] \Bigg)  \label{furier},
\end{equation}

where $E$, $p$ and $\varphi$ are respectively the energy, momentum and azimuthal angle of the particle, and $\Psi_n$ the angle of the initial state spatial plane of symmetry of the $n$-th harmonic defined by the geometrical distribution of the nucleons participating in the collision. In order to determine the second harmonic coefficient $v_{\rm 2}$, the following $\vv{Q}_{2}$ vector is measured from the azimuthal distribution of charged particles (so called ReFerence Particles RFP):

\begin{equation}
  \vv{Q}_{2} = \sum_{i = 1}^{N} w_{i} e^{2i\varphi_{i}},
\label{qvector}
\end{equation}

where $\varphi_i$ are the azimuthal angles and $N$ the multiplicity of the RFP~\cite{PhysRevC.58.1671}. The weights $w_{i}$ are described later in the text. The azimuthal angle of the $\vv{Q}_{2}$ vector

\begin{equation}
\psi_{2} = \frac{1}{2} \tan^{-1} \Bigg(\frac{Q_{2,y}}{Q_{2,x}}\Bigg),
\end{equation}

is denoted by event plane angle and is an estimate of the second harmonic symmetry plane angle $\Psi_{2}$~\cite{Voloshin:1994mz}.

The event plane (EP) and scalar product (SP) methods are used to measure the elliptic flow of inclusive electrons. The two methods are described in detail in the second part of this section. Both methods use the $\vv{Q}_{2}$ vector, which is determined with the signal amplitudes in the V0 detectors at forward and backward rapidity for the EP method and with the reconstructed tracks in the TPC at mid-rapidity for the SP method. 
In the first case, the sum in Eq.~\ref{qvector} is running over the eight azimuthal sectors of each V0 detector and  $\varphi_i$ is defined by the central azimuth of the i-th sector. The weights $w_i$ are equal to the signal amplitude in the i-th sector for the selected event, which is proportional to the number of charged particles crossing the sector. Non-uniformities in the V0 acceptance and efficiency are corrected for using the procedure described in~\cite{PhysRevC.77.034904}. Despite these corrections, a residual modulation of up to 4\% is observed in the distribution d$N_{\rm evt}$/d$\psi_{2}$ in central collisions. The effect is corrected for using additional event weights in order to make the $\psi_{2}$ distribution flat. The weights are obtained dividing the average expected number of events per each interval of the event plane distribution by the observed number of Êevents in a given event plane interval. 
In the TPC case the weights $w_{i}$ described in~\cite{Abelev:2014ipa} are used to correct for non-uniformities in the acceptance and efficiency of the TPC.
In the second case, the sum in Eq.~\ref{qvector} is running over tracks reconstructed in the TPC and selected with the following criteria: at least 70 associated space points in the TPC out of the maximum of 159, a $\chi^{2}$ per TPC point of the momentum fit in the range 0.2 $<$ $\chi^{2}$/point $<$ 4 and a transverse momentum value in the interval 0.2 $<$ $p_{\rm T}$ $<$ 5 GeV/$c$. Additionally, tracks are rejected if their distance of closest approach to the primary vertex  is larger than 3.2 cm in the $z$ direction and 2.4 cm in the ($x$,$y$) plane. In order to minimize the non-uniformities in the azimuthal acceptance, no requirement is applied on the number of ITS hits associated to tracks. 
In the case of the scalar product method, unit track weights $w_{i}$ are used in the construction of the $\vv{Q}_{2}$ vector, and possible non-uniformities in the detector are corrected with the non-uniform acceptance correction described in~\cite{Bilandzic:2010jr}.

Following ~\cite{PhysRevC.58.1671}, the electron elliptic flow can be measured with the event plane method using the following equation:


\begin{equation}
v_{\rm 2}\{\mathrm{EP}\} 
= \frac{\langle \cos[2(\varphi - \psi_{2}) ] \rangle} {R_{2}},
\label{ep}
\end{equation}

where the brackets in the numerator indicate the average over electrons with azimuthal angle $\varphi$ at mid-rapidity in all the events. The factor $R_{2}$ is the event plane resolution correction, a quantity smaller than unity that depends on the multiplicity and $v_{\rm 2}$ of the RFP. The resolution of the event plane determined with the V0 detectors is measured with the three sub-event method~\cite{Abelev:2014ipa}, namely the signals in the V0 detectors (both A and C sides) and the tracks in the positive (0 $<$ $\eta$ $<$ 0.8)  and negative \mbox{($-$0.8 $<$ $\eta$ $<$ 0)} pseudorapidity regions of the TPC. The average $R_{2}$ values in the three centrality classes used in this analysis are about 0.57 (0--10\%), 0.77 (10--20\%) and 0.78 (20--40\%). At high $p_{\rm T}$ \mbox{(8 $<$ $p_{\rm T}$ $<$ 13 GeV/$c$)}, the hadron contamination needs to be subtracted from the inclusive electron sample. In this case the $v_{\rm 2}$ of inclusive electrons is extracted from the number of electrons, $N_{\rm in}$ and $N_{\rm out}$, in two 90$^{\circ}$-wide intervals of $\Delta \varphi$ = $\varphi - \psi_{2}$: in-plane ($-\frac{\pi}{4}$ $<$ $\Delta \varphi$ $<$ $\frac{\pi}{4}$ and $\frac{3\pi}{4}$ $<$ $\Delta \varphi$ $<$ $\frac{5\pi}{4} $) and out-of-plane ($\frac{\pi}{4}$ $<$ $\Delta \varphi$ $<$ $\frac{3\pi}{4}$ and $\frac{5\pi}{4}$ $<$ $\Delta \varphi$ $<$ $\frac{7\pi}{4} $), respectively, after statistical subtraction of the hadron contamination in each of the $\Delta \varphi$ interval. 
In this case, $v_{\rm 2}\{\mathrm{EP}\}$ is  given by:

\begin{equation}   
v_{\rm 2}\{\mathrm{EP}\} = \frac{1}{R_2} \frac{\pi}{4} \frac{N_{\rm in} - N_{\rm out}}{N_{\rm in} + N_{\rm out}}.
\label{v2eqNinNout}
\end{equation}

The yield of electron candidates that do not originate from heavy-flavour hadron decays, which can be reconstructed only statistically,  is measured in $p_{\rm T}$ and $\Delta \varphi$ intervals in order to measure the elliptic flow of background electrons. The d$N$/d$\Delta \varphi$ distributions of background electrons are then fitted in each $p_{\rm T}$ interval with the following function:


\begin{equation}   
\frac{ {\rm d}N}{{\rm d}\Delta \varphi} = N_{0} \left(1 + 2\vtwob R_{2}\cos[2(\varphi-\psi_{2})] \right),
\label{v2eqdeltaphi}
\end{equation} 

where $N_{0}$ and \vtwob~are the fit parameters. 
The effect of higher harmonics on $v_{\rm 2}$ estimated with Eq.~\ref{v2eqNinNout} and ~\ref{v2eqdeltaphi}  is assumed to be negligible.


The measurement of the elliptic flow with the scalar product method~\cite{PhysRevC.66.034904, Luzum:2012da}, a two particle correlation technique, is given by:

\begin{equation}
v_{2}\{\rm SP\} = \frac{1}{2} \left( \frac{\left\langle \vv{u}_{2}^{\rm A} \cdot \frac{\vv{Q}_{2}^{\rm B}}{\it{M}^{\rm B}} \right\rangle} 
{\sqrt{\left\langle \frac{\vv{Q}_{2}^{\rm A}}{\it{M}^{\rm A}} \cdot \frac{\vv{Q}_{2}^{\rm B }}{\it{M}^{\rm B}}\right\rangle}} + 
 \frac{ \left\langle \vv{u}_{2}^{\rm B} \cdot \frac{\vv{Q}_{2}^{\rm A}}{\it{M}^{\rm A}} \right\rangle} 
{\sqrt{\left\langle \frac{\vv{Q}_{2}^{\rm A}}{\it{M}^{\rm A}} \cdot \frac{\vv{Q}_{2}^{\rm B }}{\it{M}^{\rm B}}\right\rangle}} \right),
\label{spsym}
\end{equation}

where $M^{\rm A}$ and $M^{\rm B}$ are the multiplicities and $\vv{Q}_{2}^{\rm A}$ and $\vv{Q}_{2}^{\rm B}$ are the $\vv{Q}_{2}$ vectors of two sub-events A and B, determined from TPC tracks in the positive (0 $<$ $\eta$ $<$ 0.8) and negative ($-$0.8 $<$ $\eta$ $<$ 0) pseudorapidity regions, respectively. The brackets in the numerators indicate the average over electrons with unit vector of the momentum at the primary vertex projected on the transverse plane $\vv{u}_{2}^{A}$ ($\vv{u}_{2}^{B}$) in the sub-event A (sub-event B). The sub-event procedure is applied in order to avoid auto-correlations between the electron candidates and the $\vv{Q}_{2}$ vectors, and in order to suppress non-flow contributions, like resonance decays and particles produced within jets. 


The elliptic flow measurements carried out with the event plane method could lead to ambiguous results lying between the event-averaged mean $v_{\rm 2}$ value and the root-mean-square value, as a consequence of the presence of event-by-event flow fluctuations~\cite{Luzum:2012da}. Those ambiguities are resolved using the scalar product method, that always yields to the root-mean-square value.

\subsection{Inclusive electron elliptic flow and systematic uncertainties} \label{SystUncInclElecV2}

\begin{figure}[ht!]
\centering
\includegraphics[scale=0.82]{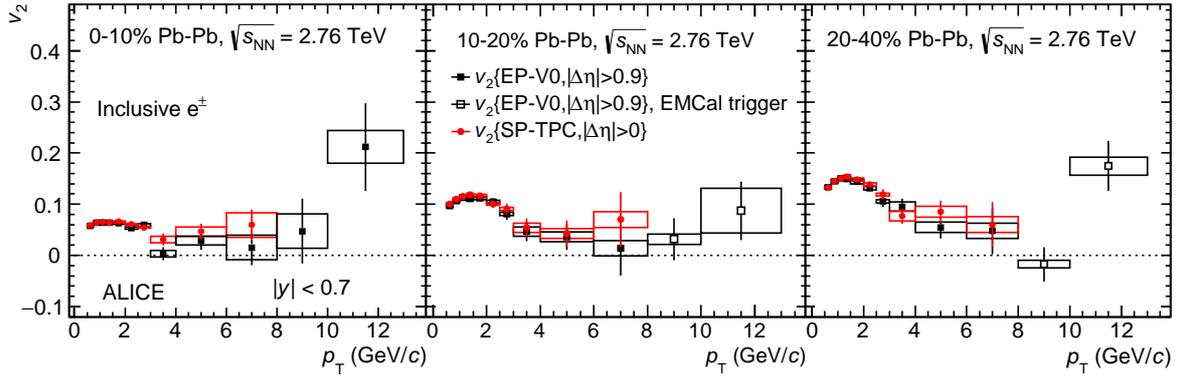}
\caption{$p_{\rm T}$-differential inclusive electron $v_{\rm 2}$ at mid-rapidity in Pb--Pb collisions at $\sqrt{s_{\rm NN}}$ = 2.76 TeV measured in the centrality classes: 0--10\% (left), 10--20\% (middle) and 20--40\% (right). The symbols are placed at the centre of the $p_{\rm T}$ interval whose width is shown by the horizontal error bars. The vertical error bars and open boxes represent the statistical and systematic uncertainties, respectively. Results with the event plane and scalar product method are reported with black and red markers, respectively. In the 10--20\% and 20--40\% centrality classes the results obtained with the EMCal trigger are reported with open black markers.}
\label{inclv2}
\end{figure}

The measured elliptic flow of inclusive electrons is shown in Figure~\ref{inclv2} in the centrality classes 0--10\%, 10--20\% and 20--40\% as a function of $p_{\rm T}$ using the event plane (black markers) and the scalar product (red markers) methods. The full markers represent the results obtained with the central and semi-central triggers, while in the 10--20\% and 20--40\% centrality classes those obtained with the EMCal trigger are reported with open markers. 
The EP and SP methods give consistent results in the full $p_{\rm T}$ region and no effects due to possible ambiguities in the EP with respect to the SP method~\cite{Luzum:2012da} are seen in this analysis. However for $p_{\rm T}$ $>$ 3 GeV/$c$ the $v_{\rm 2}$ values measured with the EP tend to be lower than those measured with the SP.
This indicates a possible stronger suppression of the non-flow effects like jet and resonance contributions with the EP method, for which the $\eta$ gap between the electron candidates and the V0 detectors is large. For both methods, the values of \vtwoi increase from central to semi-central collisions. This effect is more pronounced in the intermediate $p_{\rm T}$ region 1 $<$ $p_{\rm T}$ $<$ 4 GeV/$c$.


Several sources of systematic uncertainty affecting the electron elliptic flow measurement are considered. In the case of the EP method, two systematic uncertainty sources can affect the event plane resolution correction factor $R_{2}$. The first source arises from the presence of non-flow correlations between the two TPC sub-events used to calculate the resolution. A wider pseudorapidity gap ($|$$\Delta\eta$$|$ $>$ 0.4) is used in the systematic studies. A maximum difference of 2\% was observed in most central collisions, while in the more peripheral ones the difference was observed to be smaller than 1\%. The second contribution is due to the variation of $R_{2}$ within the centrality classes used for the analysis. The inclusive electron yield is assumed to be flat within a centrality class when computing $R_{2}$. The resulting systematic uncertainty is estimated by recomputing the $R_{2}$ value for each centrality class as weighted average of the values in finer centrality intervals (of 5 percentiles) with weights given by the corresponding electron yields. Since $R_{2}$ strongly depends on the centrality, in the most central collisions the systematic uncertainty is found to be larger (2.7\% in the 0--10\% centrality class) \mbox{than in the more peripheral ones (1\%). }

For both methods (EP and SP), the systematic uncertainty due to the hadron contamination in the electron sample is estimated for $p_{\rm T}$ $<$ 8 GeV/$c$ by comparing the inclusive electron $v_{\rm 2}$ results with the ones obtained after statistically subtracting the hadron contribution. 
The resulting uncertainty is found to be of the order of 1\% at low $p_{\rm T}$, increasing up to 5\% at $p_{\rm T}$ = 8 GeV/$c$.  

In order to study the stability of the measurements as a function of the applied selection criteria, the track selection and PID cuts are systematically varied around the value chosen in the analysis. The standard deviation of the $v_{\rm 2}$ value distribution obtained with different selection criteria in each $p_{\rm T}$ interval is taken as systematic uncertainty. This contribution is small (2\%) at low $p_{\rm T}$ ($p_{\rm T}$ $<$ 3 GeV/$c$), whereas it becomes the dominant source of uncertainty at high $p_{\rm T}$, reaching an average of 35\% over $p_{\rm T}$ and centrality class for $p_{\rm T} >$ 8 GeV/$c$ dominated by the PID cut variation.


The events selected with the EMCal trigger could have a bias in the event plane direction induced by the triggering in the limited azimuthal coverage of the EMCal detector.
According to a trigger simulation study, the effect on the elliptic flow measurement is expected to be larger for particles that do not generate a trigger signal in the detector, like hadrons, than for the particles which triggered the event (electrons, photons). The systematic uncertainty is estimated as the difference between the $v_{\rm 2}$ of charged particles in full azimuth measured in the semi-central triggered events and the $v_{\rm 2}$ of charged particles in the EMCal azimuthal coverage and triggered by the EMCal detector.
The systematic uncertainty increases with $p_{\rm T}$ and it is found to be of the order of 20\% in the 10--20\% centrality class and less than 5\% in the 20--40\% centrality class. The various systematic uncertainties are finally added in quadrature.

\subsection{Correction for background electrons} \label{NHFE}

The raw inclusive electron candidate sample consists of three main components:

\begin{enumerate}
\item electrons from heavy-flavour hadron decays and dielectron decays of quarkonia (e.g. ${\rm J}/\psi$, $\Upsilon$);
\item photonic background electrons from Dalitz decays of light neutral mesons and the conversion of their decay photons in the detector material, as well as from virtual and real thermal photons from hard scattering processes, the latter converting in the material of the detector;
\item background electrons from weak ${\rm K^{0}} \rightarrow {\rm e}^{\pm}\pi^{\mp}\nu_{e}$ (${\rm K}_{{\rm e}3}$) decays, and dielectron decays of light vector mesons.
\end{enumerate}

In this analysis, electrons from quarkonium decays are included in the definition of heavy-flavour decay electrons. The only relevant contribution arises from  ${\rm J}/\psi$ decays, which amounts to about 5.5\% in the $p_{\rm T}$ interval 3--4 GeV/$c$ in central collisions and decreases towards higher $p_{\rm T}$. 
It was estimated by using an interpolation at $\sqrt{s} = $ 2.76 TeV of the $p_{\rm T}$--differential cross section measured in pp collisions at various centre of mass energies~\cite{Bossu:2011qe} and scaling with the measured nuclear modification factor~\cite{Adam:2015rba, Chatrchyan:2012np}. 

In order to obtain the elliptic flow of heavy-flavour decay electrons, the background contributions are subtracted from the inclusive electron $v_{\rm 2}$. The background electron yield is dominated by the contribution of photonic electrons. The background from electrons from non-photonic sources, namely weak ${\rm K^{0}} \rightarrow {\rm e}^{\pm}\pi^{\mp}\nu_{e}$ (${\rm K}_{{\rm e}3}$) decays, and dielectron decays of light vector mesons, is indeed negligible as discussed in Section~\ref{cocktailsection}.
Two strategies are adopted for the electron background \vtwob~subtraction depending on $p_{\rm T}$:  
the invariant mass method~\cite{Adamczyk:2014yew} (Section~\ref{InvMassMethod}) is used
at low $p_{\rm T}$ ($p_{\rm T} <$ 1.5 GeV/$c$), while  a cocktail method~\cite{Adare:2010de} (Section~\ref{cocktailsection}) is used  for $p_{\rm T} >$ 1.5 GeV/$c$, because of the lower yield of background electrons.


\subsubsection{Invariant mass method}\label{InvMassMethod}

Electrons from direct $\gamma$ decays, $\gamma$-conversions and Dalitz-decays of $\pi^{0}$ and $\eta$ mesons are always produced in electron-positron pairs with a small invariant mass ($m_{{\rm e}^{+}{\rm e}^{-}}$) following a Kroll-Wada distribution~\cite{KrollWada:formula} peaked at zero. Such correlation does not hold for heavy-flavour decay electrons. This property is used in the invariant mass method to measure the photonic electron backgrounds. The fraction of Dalitz decays of higher mass mesons ($\omega$, $\eta'$, $\phi$), estimated with the cocktail method, is found to be negligible. Photonic electrons are reconstructed statistically by pairing an electron(positron) track with opposite charge tracks identified as positrons(electrons), called associated electrons in the following, from the same event selected with the  requirements listed in Table~\ref{table:track_cutsasso}. The pair invariant mass distribution is computed in each $p_{\rm T}$ and $\Delta \varphi$ interval of the inclusive electron tracks. The combinatorial background is subtracted using the like-sign invariant mass distribution in the same interval. 
A summary of the selection criteria applied on the electron-positron pairs is presented in Table~\ref{table:track_cutsasso}.

\vspace{0.5 cm}

\begin{table}[ht!]
\begin{center}
\centering
   \begin{tabular}{cc}
\hline
Associated electron cuts &   \\
\hline 
  $p_{\rm T}$ $^{\rm assoc}$ (GeV/$c$)& $>$ 0.15 for 0.5 $<$ $p_{\rm T}$ $<$ 3\,GeV/$c$ \\
& $>$ 0.3 for 3 $<$ $p_{\rm T}$ $<$ 8\,GeV/$c$  \\
& $>$ 0.5 for 8 $<$ $p_{\rm T}$ $<$ 13\,GeV/$c$  \\

  $|y^{\rm assoc}|$ & $<$ 0.9 \\
Number of TPC points &  $\ge$ 80\\
Number of ITS hits &  $\ge$ 2\\
DCA$^{\rm assoc}_{xy}$ & $<$ 2.4\,cm\\
DCA$^{\rm assoc}_{z}$ & $<$ 3.2\,cm \\
TPC d$E$/d$x$ cut & $-$3 $<$ $n_{\sigma}^{\rm{TPC}}$ $<$ 3 \\
\hline
Electron-positron pair cuts  &  \\
\hline
$m_{{\rm e}^{+}{\rm e}^{-}}$ (MeV/$c^2$)& $<$ 70 for 0.5 $<$ $p_{\rm T}$ $<$ 3\,GeV/$c$ \\
 & $<$ 140 for 3 $<$ $p_{\rm T}$ $<$ 13\,GeV/$c$ \\
  \hline
\end{tabular}
\caption{Selection criteria for reconstructing photonic electrons. The transverse momentum of inclusive and associated electrons is written $p_{\rm T}$ and $p_{\rm T}^{\rm assoc}$, respectively.}
\label{table:track_cutsasso}
\end{center}
\end{table}

\begin{figure}[ht!]
\centering
\includegraphics[scale=0.82]{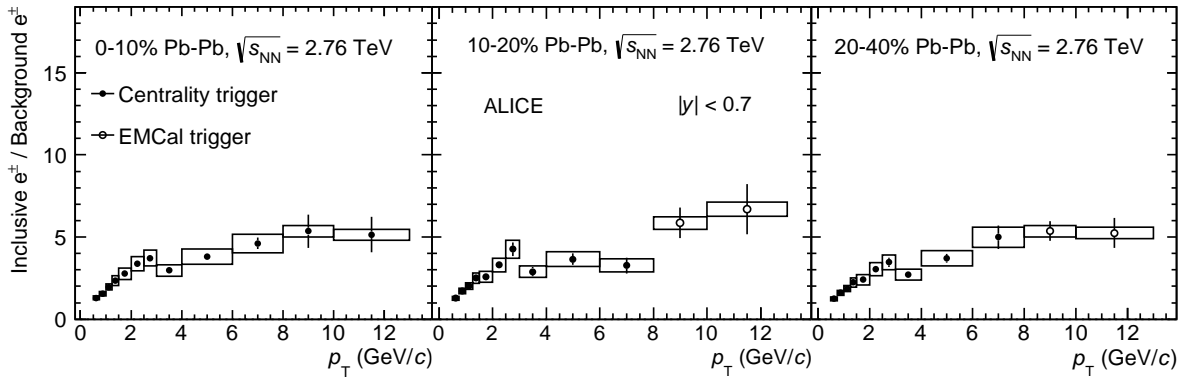}
\caption{Ratio of the inclusive electron yield to the one of background electrons obtained with the invariant mass method in Pb--Pb collisions at \energy = 2.76 TeV in \mbox{0--10\%} (left), 10--20\% (middle) and 20--40\% (right) centrality classes. The vertical error bars and open boxes represent the statistical and systematic uncertainties, respectively.}
\label{rsb}
\end{figure}

Due to detector acceptance and inefficiencies, not all photonic electrons of  the inclusive electron sample are identified with this method. Therefore, the raw yield of reconstructed photonic electrons is corrected for the efficiency to find the associated electron(positron) with the  selection criteria described above. This efficiency is estimated with Monte Carlo simulations. A sample of Pb--Pb collisions with enhanced $\pi^{0}$ and $\eta$ yields was generated with HIJING v1.36~\cite{Hijing:ref}. The transport of particles in the detector is simulated with GEANT3~\cite{Brun:1994aa}. The simulated $\pi^{0}$ and $\eta$ $p_{\rm T}$ distributions are weighted so as to match the measured $\pi^{0}$ and $\pi^{\pm}$ $p_{\rm T}$ spectra~\cite{NeutralPionSpectra,ChargedPionSpectra} and the corresponding $\eta$ $p_{\rm T}$ spectra assuming $m_{\rm T}$-scaling ~\cite{Albrecht:1995ug, Khandai:2011cf}, respectively. The photonic electron reconstruction efficiency increases with the $p_{\rm T}$ of the electron, reaching a value of about 60\% at high $p_{\rm T}$.
The inclusive-to-background ratio \mbox{(1 + $R_{SB}$)} is calculated by dividing the inclusive electron yield by the yield of photonic electrons corrected for the efficiency to find the associated electron. Figure~\ref{rsb} shows this ratio for the 0--10\% (left), 10--20\% (middle) and 20--40\% (right) centrality classes. The full markers represent the measurements obtained with the centrality-triggered samples, while in the 10--20\% and 20--40\% centrality classes the results for the EMCal-triggered sample are reported with open markers. The small decrease observed at $p_{\rm T}$ = 3 GeV/$c$ is due to the different requirements on the minimum number of hits in the SPD layers for the two electron identification strategies. For $p_{\rm T}$ larger than \mbox{2.5--3 GeV/$c$} the contribution from heavy-flavour decay electrons starts to be dominant in the inclusive electron sample.


The measurement of \vtwob~(see Eq.~\ref{hfeflow}) at low $p_{\rm T}$ ($p_{\rm T}$ $<$ 1.5 GeV/$c$) is performed with a fit to the d$N$/d$\Delta \varphi$ distributions of photonic electrons reconstructed with the invariant mass method in each $p_{\rm T}$ interval (see Eq.~\ref{v2eqdeltaphi}). 
At higher $p_{\rm T}$ \mbox{($p_{\rm T}$ $>$ 1.5\,GeV/$c$)}, the electron yield becomes too small to perform a $p_{\rm T}$ and $\Delta \varphi$-differential measurement of the photonic electrons. Figure~\ref{Cocktv2} shows the $v_{\rm 2}$ of photonic electrons measured with the invariant mass method (full markers) as a function of $p_{\rm T}$ in the centrality classes 0--10\%, 10--20\% and 20--40\%.

The systematic uncertainties of both the inclusive-to-background ratio and \vtwob\, are estimated by varying the selection criteria listed in Table~\ref{table:track_cutsasso}. For $p_{\rm T}$ $>$ 8 GeV/$c$ the TPC and EMCal PID requirements for the inclusive electron candidates are also varied in order to take into account possible systematic uncertainties from the estimation of the hadron contamination. In addition, for the inclusive-to-background ratio the small dependence of the photonic electron reconstruction efficiency on the $p_{\rm T}$ spectra of the background sources is taken into account by calculating the efficiency for different $\pi^{0}$ and $\eta$ $p_{\rm T}$ spectra. The dependence of the centrality on the systematic uncertainty of the inclusive-to-background ratio is found to be negligible. The contributions to the inclusive-to-background ratio systematic uncertainty are summarized in Table~\ref{tab:sysrsb}: the final overall systematic uncertainty is obtained summing in quadrature the different contributions. 
For \vtwob, the systematic uncertainty of the event plane correction factor $R_{2}$ is estimated using the same procedure as for the inclusive electron $v_{\rm 2}$ and is found to be the same. Moreover, the difference between the \vtwob\, measured with the invariant mass method and the one obtained with the cocktail method is taken point by point and added as an additional source of asymmetric systematic uncertainty up to $p_{\rm T}$ = 1.5 GeV/$c$  (about $-$20\% in the centrality class 0--10\% and $-$10\% in the semi-central centrality classes 10--20\%,  and 20--40\%). 
The systematic uncertainties coming from the variation of the selection criteria are found to be of the order of $\pm$20\% in the 0--10\% most-central collisions and $\pm$10 \% in the centrality classes 10--20\% and 20--40\%.
Finally, the overall systematic uncertainty on the measured \vtwob\, obtained after summing in quadrature the different contributions, are estimated to be $_{- 29\%} ^{+ 20\%}$ in the 0--10\% centrality class and $_{-15\%}^{+ 10\%}$ in the centrality classes 10--20\% and 20--40\%.

\vspace{0.5 cm}

\begin{table}[!ht]
\centering
\begin{tabular}{lccccc}
  \hline
            $p_{\rm T}$ range (GeV/$c$):  & 0.5--1.25 & 1.25--3 & 3--8 & 8--13 \\
  \hline
    Minimum number of TPC points &  2\% & 2\%  & 5\%  & --  \\
      \,\,\,\,\,for the associated electrons  &  & & & \\
    Minimum $p_{\rm T}$ of the associated electrons & 6\% & 6\% & -- & --  \\
    Maximum $m_{e^{+}e^{-}}$ & 5\% & 5\% & 10\% & 5\% \\
      \,\,\,\,\,for the electron-positron pair & & & &\\
    Influence of the $p_{\rm T}$ spectra & 5\% & 10\% & 5\% & 3\% \\
     \,\,\,\,\,of photonic sources &  &  &  & \\
    Hadron contamination in the inclusive electron sample & -- & -- & -- & 3\% \\
\hline
\end{tabular}
\caption{Systematic uncertainties of the inclusive-to-background ratio (1 + \Rsb). The centrality dependence of these systematics is found to be negligible. (see text for more details).}
\label{tab:sysrsb}
\end{table}

\subsubsection{Cocktail method}\label{cocktailsection}

The \vtwob~was also estimated using the cocktail method. A cocktail of electron spectra from background sources is calculated using a Monte Carlo event generator of hadron decays. This method requires that the momentum and elliptic flow distributions of the relevant background sources are well known.

The following electron background sources are included in the cocktail simulation:

\begin{itemize}
\item Dalitz decays of $\pi^{0}$, $\eta$, $\omega$, $\eta'$, $\phi$
\item Dielectron decays of $\eta$, $\rho^{0}$, $\omega$, $\eta'$, $\phi$
\item Conversions of decay photons from $\pi^{0}$, $\eta$, $\rho^{0}$, $\omega$, $\eta'$
\item Real and virtual conversion of prompt and thermal photons
\end{itemize}

The contribution from dielectron decays of light vector mesons is  small (below 5\% of the total background electrons considered above). For the consistency with the invariant mass method, the contributions from K$_{e3}$ and quarkonia (e.g. $J$/$\psi$ and $\Upsilon$) decays to the inclusive electron spectrum are not included in the background cocktail. The K$_{e3}$ and $\Upsilon$ contributions are not expected to be relevant in the $p_{\rm T}$ range of the analysis.  In pp collisions at $\sqrt{s}$ = 7 TeV and $\sqrt{s}$ = 2.76 TeV, the relative contribution from K$_{e3}$ decays to the electron background was observed to decrease with $p_{\rm T}$, from a maximum of 0.5\% at $p_{\rm T}$ = 0.5 GeV/$c$ for the same track requirement in the first pixel layer~\cite{Abelev:2012xe}. It is expected to stay below 1\% in Pb--Pb collisions in the $p_{\rm T}$ range considered after taking into account the different $R_{\rm AA}$ of the $\pi^{0}$~\cite{NeutralPionSpectra} and K$^{\pm}$~\cite{ChargedPionSpectra}.

Neutral pions play an important role in the cocktail. The $p_{\rm T}$ and $v_{\rm 2}$ distributions of all light scalar and vector mesons included in the cocktail are deduced from the $\pi^{0}$ spectra assuming $m_{\rm T}$~\cite{Albrecht:1995ug, Khandai:2011cf} and \mbox{$KE_{\rm T}$~\cite{Abelev:2007qg, Adare:2006ti, Adare:2012vq, Abelev:2014pua}} scaling, respectively. Indeed, electrons from $\pi^{0}$ decays are the most important background source, except in the 0--10\% and 10--20\% centrality classes for high electron $p_{\rm T}$ \mbox{($p_{\rm T}$ $>$ 8 GeV/$c$} and $p_{\rm T}$ $>$ 10 GeV/$c$, respectively), where contribution from direct photons starts to dominate. The contribution of $\pi^{0}$ decays to the electron background is twofold: via the Dalitz decay $\pi^0 \rightarrow {\rm e}^{+}{\rm e}^{-}\gamma$ and via conversions in the detector material of photons from the decay $\pi^0 \rightarrow \gamma\gamma$. 

\begin{figure}[ht!]
\begin{minipage}{18pc}
\begin{center}
\includegraphics[scale=0.4]{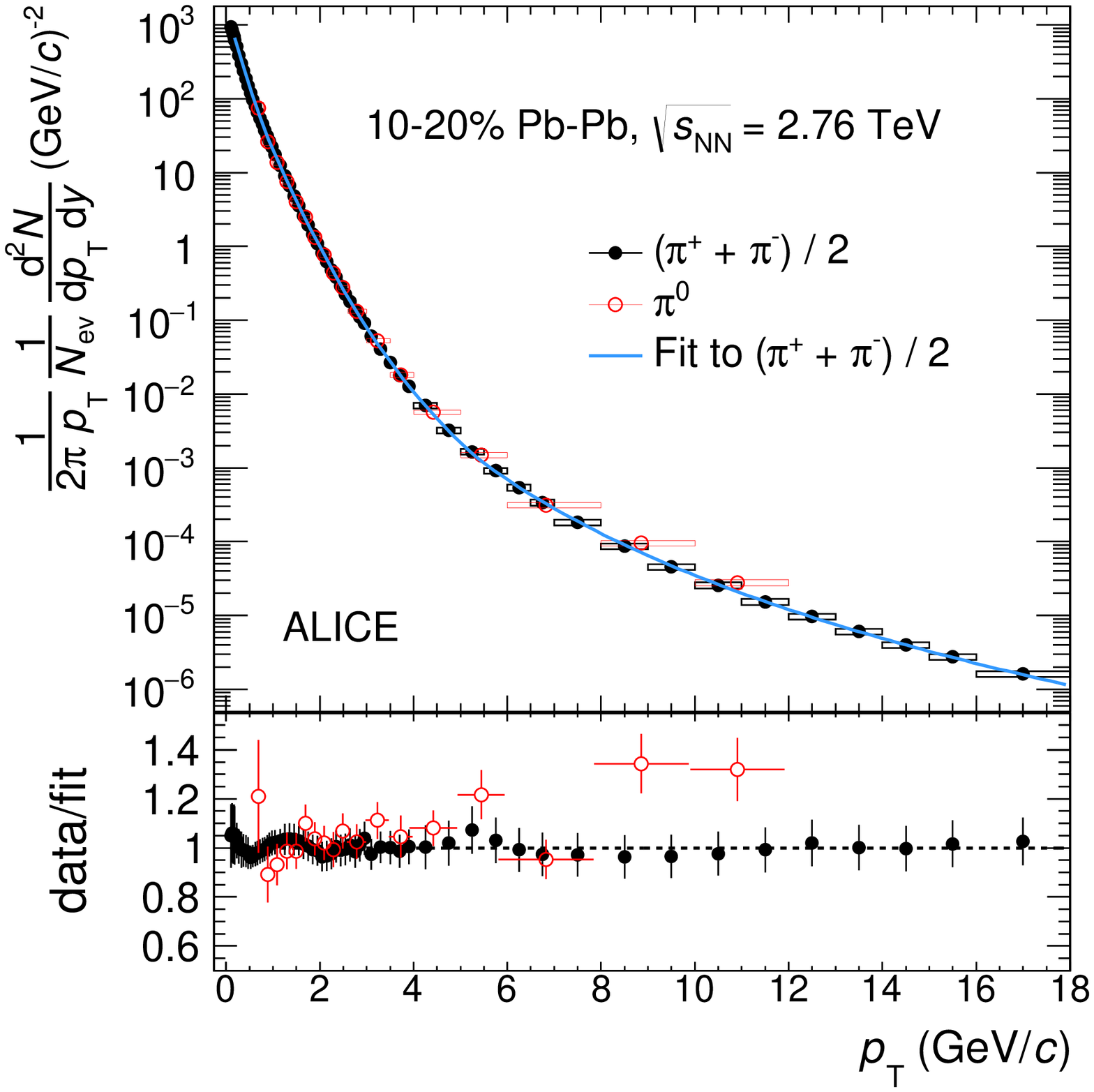}
\end{center}
\end{minipage}\hspace{1pc}%
\begin{minipage}{18pc}
\begin{center}
\includegraphics[scale=0.4]{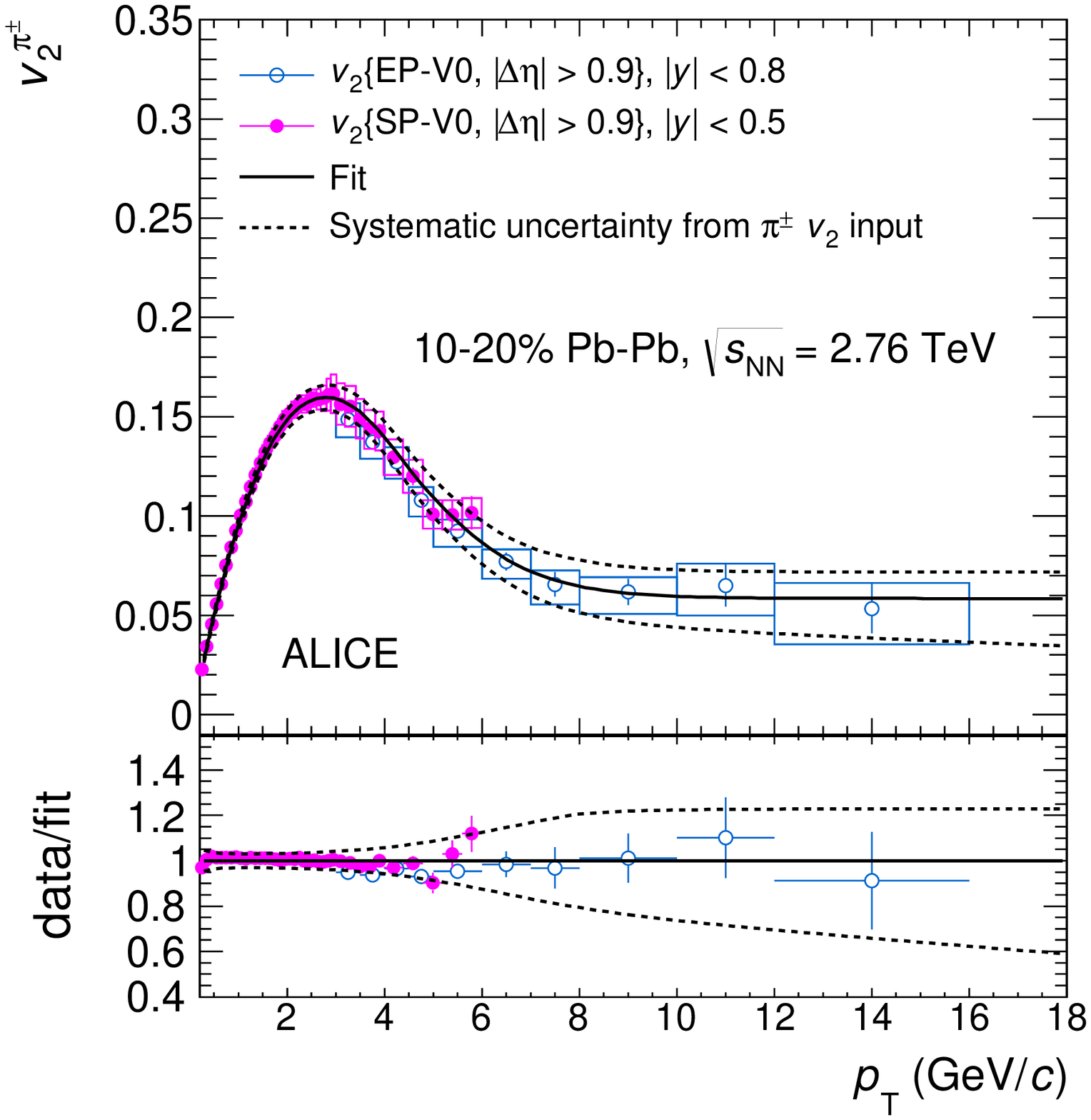}
\end{center}
\end{minipage} 
\caption{Measured $p_{\rm T}$ spectra~\cite{ChargedPionSpectra} (left) and  $v_{\rm 2}$~\cite{Abelev:2014pua,ALICEChpartv2,ALICEChpartv2note} (right)  of $\pi^{\pm}$ in the centrality class 10--20\% in Pb--Pb collisions at \energy~ = 2.76\,TeV, together with the fit and extrapolation used in the cocktail method. The ratios of data over the fit are shown on the bottom panels. The $\pi^{0}$ $p_{\rm T}$ spectrum~\cite{NeutralPionSpectra} is also shown. The vertical error bars and open boxes represent the statistical and systematic uncertainties, respectively.}
\label{fitscocktail}
\end{figure}

\begin{figure}[ht!]
\centering
\includegraphics[scale=0.4]{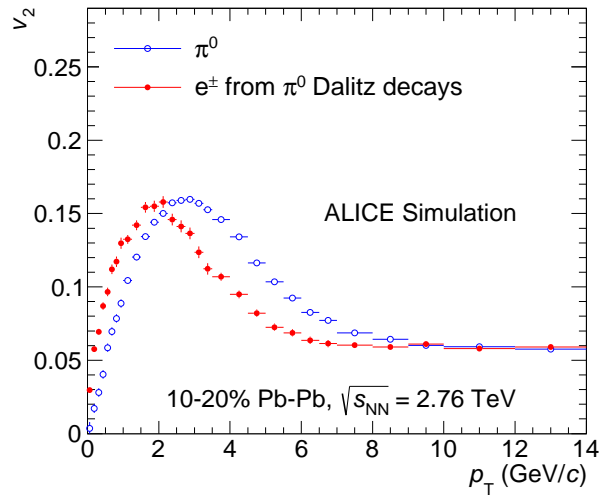}
\caption{$v_{\rm 2}$ of electrons from $\pi^{0}$ Dalitz decays (red markers) and $v_{\rm 2}$ of $\pi^{0}$ (blue markers) as a function of $p_{\rm T}$ in the centrality class 10--20\% in Pb--Pb collisions at \energy~ = 2.76 TeV as obtained from the simulation used in the cocktail method. Only statistical errors are shown.}
\label{decaycocktail}
\end{figure}

In principle, the $\pi^0$ $p_{\rm T}$ and $v_{\rm 2}$ distributions used in the Monte Carlo event generator should be based on measured $\pi^0$ spectra~\cite{NeutralPionSpectra} and $v_{\rm 2}$. 
However, because of the higher statistical precision of the combined charged pion $p_{\rm T}$ spectra~\cite{ChargedPionSpectra} and the fact that  neutral-pion and charged-pion $p_{\rm T}$ spectra are found to be consistent, 
the average of the measured charged-pion $p_{\rm T}$ spectra, $(\pi^+ + \pi^-)/2$, is used as input for the cocktail calculations. The upper-left panel of Figure~\ref{fitscocktail} shows the comparison of the neutral and charge-averaged yields of pions in the centrality class \mbox{10--20\%} together with a fit to the $\pi^{\pm}$ data with a modified Hagedorn function~\cite{2006EPJC...48..597B}. The $p_{\rm T}$ spectra are extrapolated up to 25\,GeV/$c$ using the fit function. In the last $p_{\rm T}$ interval of the measured inclusive electron spectra (10 $<$ $p_{\rm T}$ $<$ 13 GeV/$c$), about 10\% of electrons from Dalitz $\pi^{0}$ decays are expected to come from a $\pi^{0}$ with a $p_{\rm T}$ larger than 25 GeV/$c$. At such high $p_{\rm T}$, due to the similar $v_{\rm 2}$ of all particle species at high $p_{\rm T}$, this contribution is found to be negligible.
The systematic uncertainty on the heavy-flavour decay electron $v_{\rm 2}$ arising from the background sources is estimated to be smaller than 6\% in the last two $p_{\rm T}$ intervals 8--10 and 10--13 GeV/$c$. The bottom-left panel of Figure~\ref{fitscocktail} shows the ratio of the $\pi^{\pm}$ data, as well as $\pi^{0}$ data, to the fit function. The former is consistent with unity within 5\% over the full $p_{\rm T}$ range, whereas the latter is considered in the $v_{2}^{\rm Bkg}$ systematic uncertainties.

The $p_{\rm T}$-dependent $\pi^{\pm}$ elliptic flow~\cite{Abelev:2014pua,ALICEChpartv2,ALICEChpartv2note} is used as input for the cocktail calculations. The upper-right panel of Figure~\ref{fitscocktail}  shows the $v_{\rm 2}$ of charged pions measured in the \mbox{10--20\%} centrality class together with the fit function that is used in the cocktail simulations. The ratio of the data to the fit function is presented in the bottom-right panel. The function used to fit the $v_{\rm 2}$ of charged pions is an empirical function made by the convolution of trigonometric and error functions.
Measurements performed with the scalar product~ \cite{Abelev:2014pua} and event plane~\cite{ALICEChpartv2,ALICEChpartv2note} methods have been used at low-intermediate $p_{\rm T}$ \mbox{($p_{\rm T}$ $<$ 6 GeV/$c$)} and higher $p_{\rm T}$ \mbox{(3 $<$ $p_{\rm T}$ $<$ 16 GeV/$c$)}, respectively. The scalar product and event plane methods give compatible results within the uncertainties in the common $p_{\rm T}$ range 3 $<$ $p_{\rm T}$ $<$ 6 GeV/$c$. The $v_{\rm 2}$ values are extrapolated from $p_{\rm T}$ = 16\,GeV/$c$ up to $p_{\rm T}$ = 25 GeV/$c$. The elliptic flow of electrons from $\pi^{0}$ Dalitz decays is estimated from that of $\pi^{0}$ mesons using the PYTHIA 6~\cite{Sjostrand:2007gs}  event generator to simulate the Dalitz decay.
The parameterized $v_{\rm 2}$  of $\pi^{0}$ and the one of their decay electrons are shown in Figure~\ref{decaycocktail} as a function of $p_{\rm T}$.

The treatment of electrons from photon conversions in the detector material uses the GEANT4 functionality of pair production~\cite{Agostinelli2003250}. It has been implemented in the cocktail by forcing all decay photons to produce an $\rm{e}^{+}\rm{e}^{-}$ pair immediately after their creation without propagating them through the ALICE apparatus. The contribution of electrons from photon conversions is scaled according to the radiation length of the crossed material. At low $p_{\rm T}$ ($p_{\rm T}$ $<$ 3 GeV/$c$), electron tracks are required to be associated with two hits in the SPD. The effective converter thickness is estimated to be $x$/$X_{0}$ = (0.77 $\pm$ 0.07)\%, including the beam pipe, air and part of the innermost pixel layer at $y$ = 0~\cite{Abelev:2012xe}. The indicated radiation thickness is averaged over the pseudorapidity range of the analysis. At higher $p_{\rm T}$ ($p_{\rm T}$ $>$ 3\,GeV/$c$), tracks with one hit in the SPD are also used. Therefore, the material of the second pixel layer is also taken into account, leading to an effective converter thickness of $x$/$X_{0}$ = (2.15 $\pm$ 0.11)\%~\cite{Abelev:2012xe}. The results of the cocktail for photon conversion were found to be consistent within uncertainties with a full simulation test where the generated particles were propagated through the ALICE apparatus using GEANT3~\cite{Brun:118715}. The elliptic flow of electrons from the conversion of $\pi^{0}$ decay photons is found to be comparable to the one of electrons from $\pi^{0}$ Dalitz decays.  


The contributions of direct photons, thermal photons from the hot partonic and hadronic phase and photons that could be produced in the interactions of hard scattered partons with the medium, are included in the cocktail of background electrons. 
These sources can give both electrons from photon conversion in the detector material and electrons from virtual photons.
The production of real prompt photons was measured at mid-rapidity in Pb--Pb collisions in the $p_{\rm T}$ interval 0.9--14 GeV/$c$~\cite{DirectPhotonSpectrum}. The spectra are fitted and extrapolated towards lower and higher $p_{\rm T}$ (0.5 $<~p_{\rm T}~<$ 25 GeV/$c$). At intermediate-high $p_{\rm T}$ ($p_{\rm T}$ $>$ 5 GeV/$c$), the $p_{\rm T}$ spectrum of real prompt photons has been calculated with next-to-leading-order perturbative QCD calculations for pp collisions at 2.76\,TeV~\cite{DirectphotonNLO,PhysRevD.50.4436} and scaled to fit the ALICE measurements in Pb--Pb collisions~\cite{DirectPhotonSpectrum}. This assumes that the other contributions are negligible in this $p_{\rm T}$ range and that the shape of the $p_{\rm T}$ spectra of real prompt photons is not modified in heavy-ion collisions, which is justified by the experimental results. At low $p_{\rm T}$, the dominant contribution of thermal photons in the measured real direct photon $p_{\rm T}$ spectra was taken into account by adding an exponential term to the fit function. The $p_{\rm T}$ spectra of virtual photons are obtained using the Kroll-Wada function~\cite{KrollWada:formula}. The elliptic flow of real direct photons was measured in the centrality class 0--40\%~\cite{Lohner:2012ct}. To estimate the elliptic flow in the smaller centrality classes 0--10\%, 10--20\% and 20--40\%, the measurement is scaled by the ratio of the measured charged pion $v_{\rm 2}$ in the 0--40\% centrality class. Finally, the elliptic flow of virtual photons is assumed to be identical to the one of real photons.

The elliptic flow of background electrons is estimated by summing  the various background electron sources according to their relative contribution to the total background.
The main background contributions are due to $\pi^{0}$ and prompt photons. In addition, the contributions of  thermal photons (at low $p_{\rm T}$ in the \mbox{0--10\%} and \mbox{10--20\%} most  central Pb--Pb collisions) and $\eta$ are also relevant.

\begin{figure}[!ht]
\centering
\includegraphics[scale=0.82]{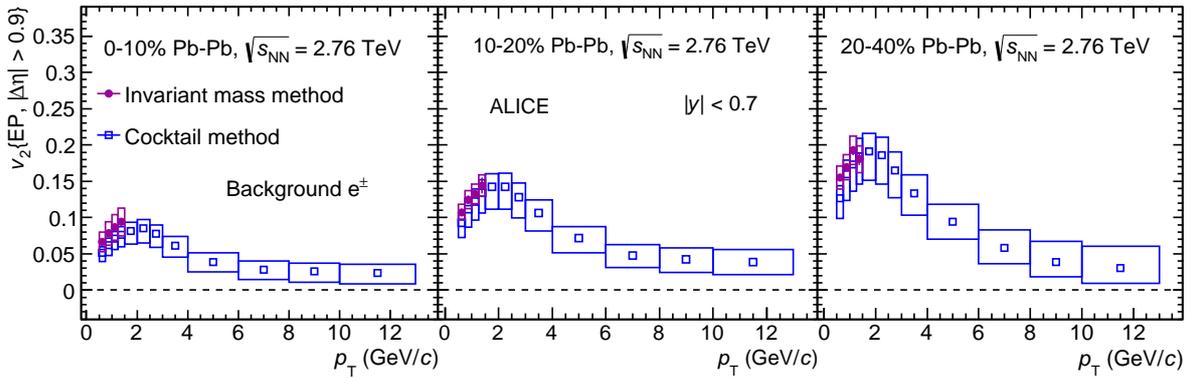}
\caption{Background electron $v_{\rm 2}$ as a function of $p_{\rm T}$ measured with the invariant mass method (full markers) and with the cocktail simulation (empty markers) in the 0--10\% (left panel), 10--20\% (middle panel) and 20--40\% (right panel) centrality classes in Pb--Pb collisions at \energy~ = 2.76 TeV.}
\label{Cocktv2}
\end{figure}

The total systematic uncertainty of \vtwob\, estimated with the cocktail method is obtained by adding in quadrature the contributions from several sources, namely:
\begin{itemize}
\item  the statistical and systematic uncertainties of the $v_{\rm 2}$ and $p_{\rm T}$ measurements of $\pi^{\pm}$ and direct photons,
\item  the quality of the fits and extrapolations of the $\pi^{\pm}$ and direct photon spectra,
\item  the systematic uncertainties on  the $KE_{\rm T}$ and $m_{\rm T}$ scaling used to estimate the $v_{\rm 2}$ and $p_{\rm T}$ distributions of higher mass mesons, respectively,
\item  the approximation of the $\pi^{0}$ $p_{\rm T}$ and $v_{\rm 2}$ distributions by the corresponding $\pi^{\pm}$ spectra.
\end{itemize}

The first one leads to the largest systematic uncertainty. It is evaluated by parameterizing the data along the upper and lower ends of their statistical and systematic uncertainties added in quadrature and generating again the complete cocktail of electron spectra based on these new parameterizations. The right panel of Figure~\ref{fitscocktail} shows examples of such fits for the $p_{\rm T}$ dependence of the $\pi^{\pm}$ $v_{\rm 2}$ in the centrality class 10--20\%. The uncertainties of the measured $p_{\rm T}$ spectra have a smaller influence on the resulting \vtwob\,\,than those of the measured $v_{\rm 2}$ spectra. 
The uncertainty on the $KE_{\rm T}$ scaling assumption is estimated by comparing the kaon $v_{\rm 2}$ obtained by $KE_{\rm T}$ scaling to the measured one~\cite{Abelev:2014pua}. The resulting systematic uncertainty is  8\% for 0--10\%, 6\% for 10--20\% and 4\% for 20--40\%. These numbers are consistent with those reported in~\cite{Abelev:2014pua}. Because of their similar mass, it is expected that the elliptic flow of $\eta$ and the one of K are similar and thus these numbers are taken directly for the $\eta$ $KE_{\rm T}$ scaling uncertainty. For the other heavier mesons the $KE_{\rm T}$ scaling does not hold precisely~\cite{Abelev:2014pua, Adare:2012vq}; however, these other particles have an extremely low weight in the cocktail, and thus these  uncertainties are neglected. The $m_{\rm T}$-scaling approach ensures that, at high $p_{\rm T}$, the transverse-momentum distributions are the same for all meson species. The normalization of the heavier meson spectra relative to the pion spectrum was determined by the ratios of heavier meson yields to neutral pion yields at high $p_{\rm T}$ ($p_{\rm T}$ $>$ 5 GeV/$c$). The values together with their uncertainties used in the analysis are taken from~\cite{Khandai:2011cf}. At low $p_{\rm T}$ ($p_{\rm T}$ $<$ 3--4 GeV/$c$) some deviations from the $m_{\rm T}$-scaling approach are expected due to in-medium effects like radial flow. 
The $m_{\rm T}$-scaling based cocktail is found to be in agreement within statistical uncertainties with a cocktail based on the $\eta$/$\pi^{0}$-ratio measured in pp collisions at $\sqrt{s}$ = 7 TeV~\cite{ALICE2012etaOverPi}.
Also, due to the similarity of the elliptic flow of decay electrons and conversion electrons originating from the dominating mother mesons ($\pi^{0}$ and $\eta$), the material budget uncertainty was found to have no significant effect.


Two additional sources of systematic uncertainty related to the electron track reconstruction were studied. First, reconstructed electron candidates have a limited $p_{\rm T}$ resolution. In particular, Bremsstrahlung in the detector material shifts their reconstructed $p_{\rm T}$ towards lower values. Secondly, hits in the SPD can be wrongly associated to a track with a probability increasing with decreasing $p_{\rm T}$. This leads to an increase of the amount of electrons from photon conversions occurring beyond the SPD layers in the inclusive electron sample and a degradation of the $p_{\rm T}$ and $\varphi$ resolutions of tracks used in the analysis. The resulting effects on $v_{2}^{\rm Bkg}$ were evaluated with the cocktail method using SPD hit mismatch probabilities and resolution maps obtained with a full simulation of the ALICE apparatus. No significant change of \vtwob\,\,was observed .

The \vtwob\,\, estimated with the cocktail method is shown as a function of $p_{\rm T}$ \mbox{(0.5 $<$ $p_{\rm T}$ $<$ 13 GeV/$c$)} in the centrality classes 0--10\%, 10--20\% and 20--40\% in Figure~\ref{Cocktv2}, together with the one obtained with the invariant mass method (0.5 $<$ $p_{\rm T}$ $<$ 1.5 GeV/$c$). The results are consistent within the systematic uncertainties in  the three centrality classes.

%% file: Results.tex

\section{Results} \label{results}


The elliptic flow of heavy-flavour decay electrons \vtwohfe~is computed using Eq.~\ref{hfeflow}. The systematic uncertainties on \vtwoi, \Rsb~and \vtwob~are propagated to \vtwohfe.
The error propagation for the background subtraction is based on an approximation of a second order error propagation~\cite{D'Agostini:2000ra, 2004physics...3086D}, where differently from the Gaussian approximation, not only linear effects of the error propagation are considered but also quadratic effects. This is necessary especially in case the non-linearity of the subtraction can not be neglected anymore. The basic concept is that the upper and lower systematic errors are both found by independently varying the uncertainties of the input variables by one sigma up and down.
The value of \vtwohfe~ is obtained only with the event plane method, because the charged-pion $v_{\rm 2}$ measurements with the scalar product method are not available at high $p_{\rm T}$ for the estimation of \vtwob~using the cocktail method. At low-intermediate $p_{\rm T}$ ($p_{\rm T}$ $<$ 6 GeV/c), the \vtwohfe~extracted with the EP and the SP methods are expected to be compatible  within uncertainties, as seen from the measured inclusive electron and charged pion $v_{\rm 2}$. 

\begin{figure}[ht]
\centering
\includegraphics[scale=0.82]{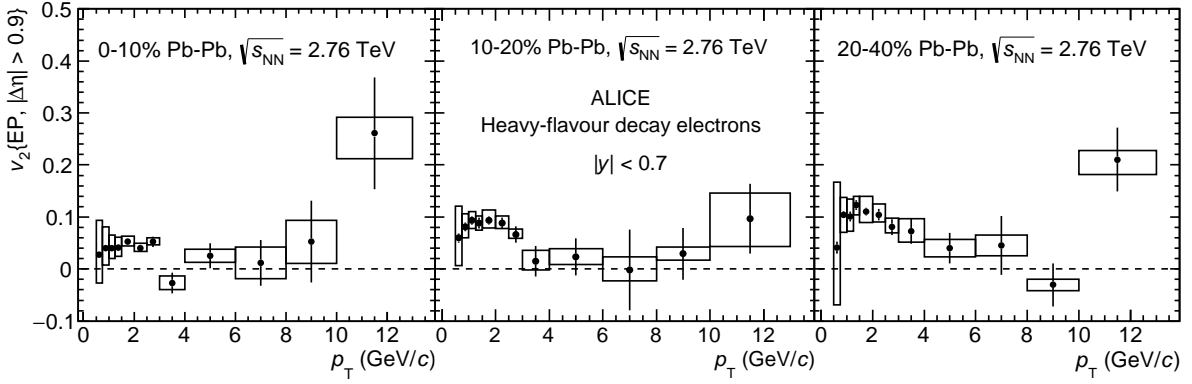}
\caption{Elliptic flow of electrons from heavy-flavour hadron decays in the 0--10\% (left panel), 10--20\% (middle panel) and 20--40\% (right panel) centrality classes in Pb--Pb collisions at \energy~ = 2.76 TeV at mid-rapidity as function of $p_{\rm T}$. The symbols are placed at the centre of the $p_{\rm T}$ interval whose width is shown by the horizontal error bar. The vertical error bars and open boxes represent the statistical and systematic uncertainties, respectively. The results are obtained with the event plane method and an eta gap $|\Delta \eta|$ $>$ 0.9.}
\label{HFEv2}
\end{figure}

Figure~\ref{HFEv2} shows the elliptic flow of electrons from heavy-flavour hadron decays at mid-rapidity ($|$$y$$|$ $<$ 0.7) as a function of $p_{\rm T}$ in  Pb--Pb collisions at \energy~ = 2.76 TeV for the 0--10\%, 10--20\% and 20--40\% centrality classes. At low $p_{\rm T}$, the systematic uncertainties are large because of the small signal-to-background ratio. The central value of \vtwohfe~is slightly increasing with $p_{\rm T}$ up to $\sim$ 1.5 GeV/$c$ where it reaches a maximum in all centrality classes. A positive $v_{\rm 2}$ is observed in all centrality classes, with a maximum significance of 5.9$\sigma$ in the $p_{\rm T}$ interval \mbox{2--2.5 GeV/$c$} in semi-central collisions (20--40\%). At higher $p_{\rm T}$, the measured $v_{\rm 2}$ of heavy-flavour decay electrons exhibits a slight decrease as $p_{\rm T}$ increases, becoming consistent with zero within large uncertainties for $p_{\rm T}$ $>$ 4 GeV/$c$. A positive $v_{\rm 2}$ is also observed in the $p_{\rm T}$ interval \mbox{10--13 GeV/$c$} in the 20--40\% centrality class, however the large uncertainties do not allow for a conclusion.

\begin{figure}[h]
\centering
\includegraphics[scale=0.37]{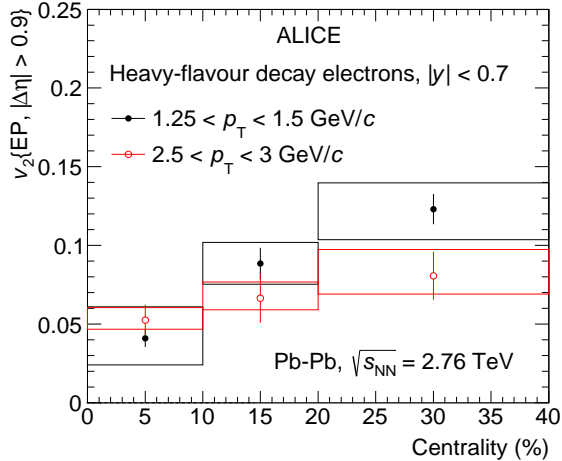}
\caption{Elliptic flow of electrons from heavy-flavour hadron decays at mid-rapidity as a function of the centrality class in Pb--Pb collisions at \energy~ = 2.76 TeV. The symbols are placed at the centre of the centrality interval whose width is shown by the horizontal error bar. The vertical error bars and open boxes represent the statistical and systematic uncertainties, respectively.}
\label{HFEv2int}
\end{figure}

\begin{figure}[h]
\centering
\includegraphics[scale=0.82]{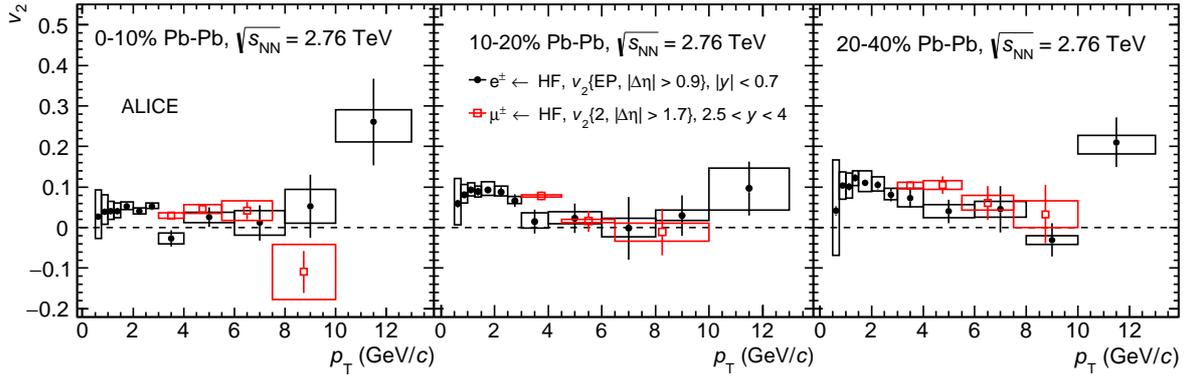}
\caption{Elliptic flow of heavy-flavour decay electrons at mid-rapidity ($|y|$ $<$ 0.7) (closed symbols) as a function of $p_{\rm T}$ compared to the elliptic flow of heavy-flavour decay muons at forward rapidity ~\cite{Adam:2015pga} (2.5 $<$ $y$ $<$ 4) (open symbols) in the 0--10\% (left panel), 10--20\% (middle panel) and 20--40\% (right panel) centrality classes in Pb--Pb collisions at \energy~ = 2.76 TeV. The symbols are placed at the centre of the $p_{\rm T}$ interval whose width is shown by the horizontal error bar. The vertical error bars and open boxes represent the statistical and systematic uncertainties, respectively.}
\label{HFEv2emu}
\end{figure}

Figure~\ref{HFEv2int} shows the centrality dependence of the elliptic flow of heavy-flavour decay electrons in two $p_{\rm T}$ intervals (1.25--1.5 GeV/$c$ and 2.5--3 GeV/$c$).  In the interval 1.25 $<$ $p_{\rm T}$ $<$ 1.5 GeV/$c$ the contribution from charm hadron decays is expected to be dominant in the heavy-flavour decay electron sample, whereas in the higher $p_{\rm T}$ interval the beauty-hadron decays should start to be relevant. In pp collisions at $\sqrt{s}$ = 2.76 TeV, beauty hadron decays are indeed the dominant source of heavy-flavour decay electrons for $p_{\rm T}$ $>$ 4.5 GeV/$c$~\cite{hfepp276}. A decreasing trend of \vtwohfe\,\,towards central collisions is observed. This is consistent with a final-state anisotropy in momentum space driven by the initial geometrical anisotropy of the nucleons participating in the collision, which increases towards peripheral collisions. 
This result indicates that the interactions with the medium constituents transfer to heavy quarks, mainly charm, information on the azimuthal anisotropy of the system, possibly suggesting that charm quarks participate in the collective expansion of the system.



The elliptic flow of prompt D mesons was measured at mid-rapidity in the centrality classes 0--10\%, 10--30\% and 30--50\% for $p_{\rm T}$ $>$ 2 GeV/$c$~\cite{Abelev:2013lca,Abelev:2014ipa}. The results are similar to those of heavy-flavour decay electrons after taking into account the decay kinematics, which shifts their maximum value of $v_{\rm 2}$ to lower $p_{\rm T}$ with respect to their parent D mesons. 
At forward rapidity (2.5 $<$ $y$ $<$ 4), the elliptic flow of heavy-flavour decay muons \vtwohfm was measured with various methods in the centrality classes 0--10\%, 10--20\% and 20--40\%~\cite{Adam:2015pga}. Figure~\ref{HFEv2emu} shows the comparison of \vtwohfe at mid-rapidity and \vtwohfm at foward rapidity obtained with the two-particle $Q$-cumulant method with $|\Delta\eta|$ $>$ 1.7. The observed $v_{\rm 2}$ of heavy-flavour decay leptons is similar at mid- and forward rapidity.

\section{Comparison with model calculations} \label{modelcomparison}



\begin{figure}[!ht]
\centering
\includegraphics[scale=0.4]{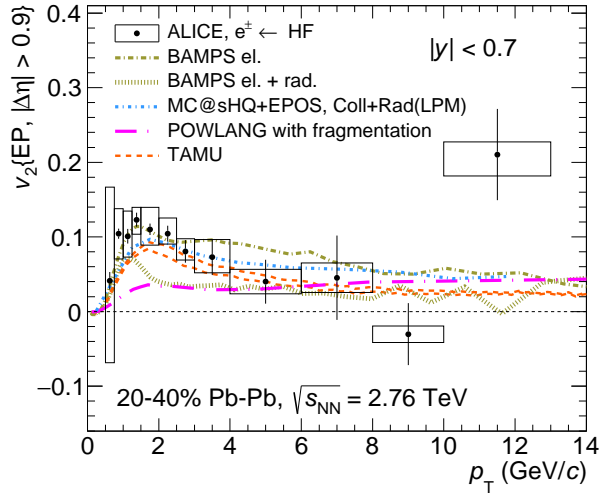}
\caption{Heavy-flavour decay electron $v_2$ at mid-rapidity as a function of $p_{\rm T}$  in semi-central Pb--Pb collisions at \energy~ = 2.76\,TeV compared to model calculations~\cite{Uphoff:2012gb,Uphoff:2013soa,He:2014cla,Alberico:2013bza,Nahrgang:2013xaa}.}
    \label{HFEv2MODE}
\end{figure}

Figure~\ref{HFEv2MODE} shows the comparison of the measured heavy-flavour decay electron elliptic flow in the \mbox{20--40\%} centrality class with theoretical model calculations. BAMPS~\cite{Uphoff:2012gb,Uphoff:2013soa} is a partonic transport model based on the Boltzmann approach to multi-parton scatterings. Two versions are presented. In the first one, BAMPS el.~\cite{Uphoff:2012gb}, heavy quarks interact with the medium via collisional (elastic) processes computed with running strong coupling constant. The binary cross section is scaled with a correction factor in order to mimic the contribution of radiative processes, which are not included. The heavy-flavour decay electron elliptic flow and nuclear modification factor measured at RHIC are used to tune this factor. In the second version, {\mbox{BAMPS el. + rad.~\cite{Uphoff:2013soa}}}, radiative processes are included as well.  In both approaches, the hadronisation uses a vacuum fragmentation function. TAMU~\cite{He:2014cla} is a heavy-flavour transport model that incorporates energy loss via collisional processes with resonance formation and dissociation in an evolving hydrodynamic medium. The hydrodynamical expansion of the medium is constrained by the measured $p_{\rm T}$ and $v_{\rm 2}$ spectra of light-flavour hadrons. The hadronisation contains a component of recombination of heavy quarks with light-flavour quarks from the QGP. Diffusion processes in the hadronic phase are also included. POWLANG~\cite{Alberico:2013bza} is a transport model based on the Langevin transport equation with collisional energy loss in an expanding, deconfined medium. Hadronisation uses a vacuum fragmentation function. A more recent version of POWLANG~\cite{2015EPJC...75..121B} uses an in-medium hadronisation resulting in a larger $v_{\rm 2}$ for the D meson. 
MC@sHQ+EPOS \cite{Nahrgang:2013xaa} is a perturbative QCD model which includes radiative (with Landau-Pomeranchuk-Migdal correction~\cite{Baier:2000mf}) and collisional energy loss in an expanding medium. A component of recombination of heavy quarks with light-flavour quarks from the QGP is also incorporated in the model. The medium fluid dynamical expansion is based \mbox{on the EPOS model~\cite{Werner:2012xh}.} 

The elliptic flow of heavy-flavour decay electrons is qualitatively described by the models including significant interactions of heavy quarks with a hydrodynamically-expanding QGP. Mechanisms like collisional processes and hadronisation via recombination transfer to heavy quarks and heavy-flavour hadrons the elliptic flow induced during the system expansion, and are able to describe  the measured positive \vtwohfe~at intermediate $p_{\rm T}$. The $p_{\rm T}$ dependence of $v_{\rm 2}$ reflects the interplay between significant scatterings with the constituents of an expanding medium at low and intermediate $p_{\rm T}$, and the path-length dependence of the parton energy loss in the hot and dense matter at high $p_{\rm T}$. Models which underestimate the elliptic flow of heavy-flavour decay electrons at low and intermediate $p_{\rm T}$ (POWLANG and {\mbox{BAMPS el. + rad}}) underestimate as well the elliptic flow of prompt D mesons at mid-rapidity~\cite{Abelev:2014ipa, Andronic:2015wma}. 
Similarly BAMPS el. which reproduces qualitatively the elliptic flow of heavy-flavour decay electrons, describes at mid-rapidity the prompt D meson $v_{\rm 2}$~\cite{Abelev:2014ipa} and at forward rapidity the heavy-flavour decay muon $v_{\rm 2}$~\cite{Adam:2015pga}.

%% file: Conclusion.tex

\section{Conclusions} \label{section6}

We presented the elliptic flow of electrons from heavy-flavour hadron decays at mid-rapidity ($|y|$ $<$ 0.7) in central and semi-central Pb--Pb collisions at \energy~ = 2.76 TeV measured with ALICE at the LHC. The results are presented as a function of the transverse momentum in the interval 0.5 $<$ $p_{\rm T}$ $<$ 13 GeV/$c$ in three centrality classes (0--10\%, 10--20\%, and 20--40\%). The $p_{\rm T}$ dependence of the heavy-flavour decay electron $v_{\rm 2}$ shows  a positive \vtwo\,\,at low and intermediate $p_{\rm T}$ in all centrality classes with a significance of 5.9$\sigma$ in the $p_{\rm T}$ range 2 $<$ $p_{\rm T}$ $<$ 2.5 GeV/$c$ in semi-central  (20--40\%) collisions. This result indicates that the interactions with the medium constituents transfer to heavy quarks, mainly charm, information on the azimuthal anisotropy of the system, possibly suggesting that charm quarks participate in the collective expansion of the system. At higher $p_{\rm T}$ ($p_{\rm T}$ $>$ 4 GeV/$c$) the measured $v_{\rm 2}$ is consistent with zero within large uncertainties. The centrality dependence of the heavy-flavour decay electron elliptic flow was studied in two $p_{\rm T}$ intervals (1.25--1.5 GeV/$c$ and 2.5--3 GeV/$c$).  At low $p_{\rm T}$ the contribution from charm hadron decays is expected to be dominant, whereas it decreases at higher $p_{\rm T}$.  A decrease of $v_{\rm 2}$ of electrons from heavy-flavour hadron decays  towards more central collisions is observed in particular at low transverse momentum (1.25 $<$ $p_{\rm T}$ $<$ 1.5 GeV/$c$). Such a trend is expected from the increase of the initial spatial anisotropy from central to peripheral collisions. The elliptic flow of heavy-flavour decay electrons at mid-rapidity ($|y|$ $<$ 0.7) is found to be similar to the one of heavy-flavour decay muons at forward rapidity (2.5 $<$ y $<$ 4). The elliptic flow of electrons from heavy-flavour hadron decays is compared to theoretical model calculations. The anisotropy is best described by models that include significant interactions of heavy quarks with the medium and mechanisms, like collisional energy loss and hadronisation via recombination, that transfer to heavy quarks and heavy-flavour hadrons the elliptic flow produced  during the system expansion. 

%% file: acknowledgements.tex

The ALICE Collaboration would like to thank all its engineers and technicians for their invaluable contributions to the construction of the experiment and the CERN accelerator teams for the outstanding performance of the LHC complex.
The ALICE Collaboration gratefully acknowledges the resources and support provided by all Grid centres and the Worldwide LHC Computing Grid (WLCG) collaboration.
The ALICE Collaboration acknowledges the following funding agencies for their support in building and
running the ALICE detector:
State Committee of Science,  World Federation of Scientists (WFS)
and Swiss Fonds Kidagan, Armenia;
Conselho Nacional de Desenvolvimento Cient\'{\i}fico e Tecnol\'{o}gico (CNPq), Financiadora de Estudos e Projetos (FINEP),
Funda\c{c}\~{a}o de Amparo \`{a} Pesquisa do Estado de S\~{a}o Paulo (FAPESP);
Ministry of Science \& Technology of China (MSTC), National Natural Science Foundation of China (NSFC) and Ministry of Education of China (MOEC)";
Ministry of Science, Education and Sports of Croatia and  Unity through Knowledge Fund, Croatia;
Ministry of Education and Youth of the Czech Republic;
Danish Natural Science Research Council, the Carlsberg Foundation and the Danish National Research Foundation;
The European Research Council under the European Community's Seventh Framework Programme;
Helsinki Institute of Physics and the Academy of Finland;
French CNRS-IN2P3, the `Region Pays de Loire', `Region Alsace', `Region Auvergne' and CEA, France;
German Bundesministerium fur Bildung, Wissenschaft, Forschung und Technologie (BMBF) and the Helmholtz Association;
General Secretariat for Research and Technology, Ministry of Development, Greece;
National Research, Development and Innovation Office (NKFIH), Hungary;
Council of Scientific and Industrial Research (CSIR), New Delhi;
Department of Atomic Energy and Department of Science and Technology of the Government of India;
Istituto Nazionale di Fisica Nucleare (INFN) and Centro Fermi - Museo Storico della Fisica e Centro Studi e Ricerche ``Enrico Fermi'', Italy;
Japan Society for the Promotion of Science (JSPS) KAKENHI and MEXT, Japan;
National Research Foundation of Korea (NRF);
Consejo Nacional de Cienca y Tecnologia (CONACYT), Direccion General de Asuntos del Personal Academico(DGAPA), M\'{e}xico, Amerique Latine Formation academique - 
European Commission~(ALFA-EC) and the EPLANET Program~(European Particle Physics Latin American Network);
Stichting voor Fundamenteel Onderzoek der Materie (FOM) and the Nederlandse Organisatie voor Wetenschappelijk Onderzoek (NWO), Netherlands;
Research Council of Norway (NFR);
Pontificia Universidad Cat\'{o}lica del Per\'{u};
National Science Centre, Poland;
Ministry of National Education/Institute for Atomic Physics and National Council of Scientific Research in Higher Education~(CNCSI-UEFISCDI), Romania;
Joint Institute for Nuclear Research, Dubna;
Ministry of Education and Science of Russian Federation, Russian Academy of Sciences, Russian Federal Agency of Atomic Energy, Russian Federal Agency for Science and Innovations and The Russian Foundation for Basic Research;
Ministry of Education of Slovakia;
Department of Science and Technology, South Africa;
Centro de Investigaciones Energeticas, Medioambientales y Tecnologicas (CIEMAT), E-Infrastructure shared between Europe and Latin America (EELA), 
Ministerio de Econom\'{i}a y Competitividad (MINECO) of Spain, Xunta de Galicia (Conseller\'{\i}a de Educaci\'{o}n),
Centro de Aplicaciones Tecnológicas y Desarrollo Nuclear (CEA\-DEN), Cubaenerg\'{\i}a, Cuba, and IAEA (International Atomic Energy Agency);
Swedish Research Council (VR) and Knut $\&$ Alice Wallenberg Foundation (KAW);
National Science and Technology Development Agency (NSDTA), Suranaree University of Technology (SUT) and Office of the Higher Education Commission under NRU project of Thailand;
Ukraine Ministry of Education and Science;
United Kingdom Science and Technology Facilities Council (STFC);
The United States Department of Energy, the United States National Science Foundation, the State of Texas, and the State of Ohio.

%% file: Alice_Authorlist_2016-05-05.tex


\begingroup
\small
\begin{flushleft}
J.~Adam$^\textrm{\scriptsize 39}$,
D.~Adamov\'{a}$^\textrm{\scriptsize 85}$,
M.M.~Aggarwal$^\textrm{\scriptsize 89}$,
G.~Aglieri Rinella$^\textrm{\scriptsize 35}$,
M.~Agnello$^\textrm{\scriptsize 112}$\textsuperscript{,}$^\textrm{\scriptsize 31}$,
N.~Agrawal$^\textrm{\scriptsize 48}$,
Z.~Ahammed$^\textrm{\scriptsize 136}$,
S.~Ahmad$^\textrm{\scriptsize 18}$,
S.U.~Ahn$^\textrm{\scriptsize 69}$,
S.~Aiola$^\textrm{\scriptsize 140}$
% ...
\renewcommand\labelenumi{\textsuperscript{\theenumi}~}

\section*{Affiliation notes}
\renewcommand\theenumi{\roman{enumi}}
\begin{Authlist}
\item \Adef{0}Deceased
\item \Adef{idp1821440}{Also at: Georgia State University, Atlanta, Georgia, United States}
\item \Adef{idp3238368}{Also at: Also at Department of Applied Physics, Aligarh Muslim University, Aligarh, India}
\item \Adef{idp3964480}{Also at: M.V. Lomonosov Moscow State University, D.V. Skobeltsyn Institute of Nuclear, Physics, Moscow, Russia}
\end{Authlist}

\section*{Collaboration Institutes}
\renewcommand\theenumi{\arabic{enumi}~}

$^{1}$A.I. Alikhanyan National Science Laboratory (Yerevan Physics Institute) Foundation, Yerevan, Armenia
\\
$^{2}$Benem\'{e}rita Universidad Aut\'{o}noma de Puebla, Puebla, Mexico
\\
$^{3}$Bogolyubov Institute for Theoretical Physics, Kiev, Ukraine
\\
$^{4}$Bose Institute, Department of Physics 
and Centre for Astroparticle Physics and Space Science (CAPSS), Kolkata, India
\\
$^{5}$Budker Institute for Nuclear Physics, Novosibirsk, Russia
\\
$^{6}$California Polytechnic State University, San Luis Obispo, California, United States
\\
$^{7}$Central China Normal University, Wuhan, China
\\
$^{8}$Centre de Calcul de l'IN2P3, Villeurbanne, Lyon, France
\\
$^{9}$Centro de Aplicaciones Tecnol\'{o}gicas y Desarrollo Nuclear (CEADEN), Havana, Cuba
\\
$^{10}$Centro de Investigaciones Energ\'{e}ticas Medioambientales y Tecnol\'{o}gicas (CIEMAT), Madrid, Spain
\\
$^{11}$Centro de Investigaci\'{o}n y de Estudios Avanzados (CINVESTAV), Mexico City and M\'{e}rida, Mexico
\\
$^{12}$Centro Fermi - Museo Storico della Fisica e Centro Studi e Ricerche ``Enrico Fermi', Rome, Italy
\\
$^{13}$Chicago State University, Chicago, Illinois, United States
\\
$^{14}$China Institute of Atomic Energy, Beijing, China
\\
$^{15}$Commissariat \`{a} l'Energie Atomique, IRFU, Saclay, France
\\
$^{16}$COMSATS Institute of Information Technology (CIIT), Islamabad, Pakistan
\\
$^{17}$Departamento de F\'{\i}sica de Part\'{\i}culas and IGFAE, Universidad de Santiago de Compostela, Santiago de Compostela, Spain
\\
$^{18}$Department of Physics, Aligarh Muslim University, Aligarh, India
\\
$^{19}$Department of Physics, Ohio State University, Columbus, Ohio, United States
\\
$^{20}$Department of Physics, Sejong University, Seoul, South Korea
\\
$^{21}$Department of Physics, University of Oslo, Oslo, Norway
\\
$^{22}$Department of Physics and Technology, University of Bergen, Bergen, Norway
\\
$^{23}$Dipartimento di Fisica dell'Universit\`{a} 'La Sapienza'
and Sezione INFN, Rome, Italy
\\
$^{24}$Dipartimento di Fisica dell'Universit\`{a}
and Sezione INFN, Cagliari, Italy
\\
$^{25}$Dipartimento di Fisica dell'Universit\`{a}
and Sezione INFN, Trieste, Italy
\\
$^{26}$Dipartimento di Fisica dell'Universit\`{a}
and Sezione INFN, Turin, Italy
\\
$^{27}$Dipartimento di Fisica e Astronomia dell'Universit\`{a}
and Sezione INFN, Bologna, Italy
\\
$^{28}$Dipartimento di Fisica e Astronomia dell'Universit\`{a}
and Sezione INFN, Catania, Italy
\\
$^{29}$Dipartimento di Fisica e Astronomia dell'Universit\`{a}
and Sezione INFN, Padova, Italy
\\
$^{30}$Dipartimento di Fisica `E.R.~Caianiello' dell'Universit\`{a}
and Gruppo Collegato INFN, Salerno, Italy
\\
$^{31}$Dipartimento DISAT del Politecnico and Sezione INFN, Turin, Italy
\\
$^{32}$Dipartimento di Scienze e Innovazione Tecnologica dell'Universit\`{a} del Piemonte Orientale and INFN Sezione di Torino, Alessandria, Italy
\\
$^{33}$Dipartimento Interateneo di Fisica `M.~Merlin'
and Sezione INFN, Bari, Italy
\\
$^{34}$Division of Experimental High Energy Physics, University of Lund, Lund, Sweden
\\
$^{35}$European Organization for Nuclear Research (CERN), Geneva, Switzerland
\\
$^{36}$Excellence Cluster Universe, Technische Universit\"{a}t M\"{u}nchen, Munich, Germany
\\
$^{37}$Faculty of Engineering, Bergen University College, Bergen, Norway
\\
$^{38}$Faculty of Mathematics, Physics and Informatics, Comenius University, Bratislava, Slovakia
\\
$^{39}$Faculty of Nuclear Sciences and Physical Engineering, Czech Technical University in Prague, Prague, Czech Republic
\\
$^{40}$Faculty of Science, P.J.~\v{S}af\'{a}rik University, Ko\v{s}ice, Slovakia
\\
$^{41}$Faculty of Technology, Buskerud and Vestfold University College, Tonsberg, Norway
\\
$^{42}$Frankfurt Institute for Advanced Studies, Johann Wolfgang Goethe-Universit\"{a}t Frankfurt, Frankfurt, Germany
\\
$^{43}$Gangneung-Wonju National University, Gangneung, South Korea
\\
$^{44}$Gauhati University, Department of Physics, Guwahati, India
\\
$^{45}$Helmholtz-Institut f\"{u}r Strahlen- und Kernphysik, Rheinische Friedrich-Wilhelms-Universit\"{a}t Bonn, Bonn, Germany
\\
$^{46}$Helsinki Institute of Physics (HIP), Helsinki, Finland
\\
$^{47}$Hiroshima University, Hiroshima, Japan
\\
$^{48}$Indian Institute of Technology Bombay (IIT), Mumbai, India
\\
$^{49}$Indian Institute of Technology Indore, Indore, India
\\
$^{50}$Indonesian Institute of Sciences, Jakarta, Indonesia
\\
$^{51}$Inha University, Incheon, South Korea
\\
$^{52}$Institut de Physique Nucl\'eaire d'Orsay (IPNO), Universit\'e Paris-Sud, CNRS-IN2P3, Orsay, France
\\
$^{53}$Institute for Nuclear Research, Academy of Sciences, Moscow, Russia
\\
$^{54}$Institute for Subatomic Physics of Utrecht University, Utrecht, Netherlands
\\
$^{55}$Institute for Theoretical and Experimental Physics, Moscow, Russia
\\
$^{56}$Institute of Experimental Physics, Slovak Academy of Sciences, Ko\v{s}ice, Slovakia
\\
$^{57}$Institute of Physics, Academy of Sciences of the Czech Republic, Prague, Czech Republic
\\
$^{58}$Institute of Physics, Bhubaneswar, India
\\
$^{59}$Institute of Space Science (ISS), Bucharest, Romania
\\
$^{60}$Institut f\"{u}r Informatik, Johann Wolfgang Goethe-Universit\"{a}t Frankfurt, Frankfurt, Germany
\\
$^{61}$Institut f\"{u}r Kernphysik, Johann Wolfgang Goethe-Universit\"{a}t Frankfurt, Frankfurt, Germany
\\
$^{62}$Institut f\"{u}r Kernphysik, Westf\"{a}lische Wilhelms-Universit\"{a}t M\"{u}nster, M\"{u}nster, Germany
\\
$^{63}$Instituto de Ciencias Nucleares, Universidad Nacional Aut\'{o}noma de M\'{e}xico, Mexico City, Mexico
\\
$^{64}$Instituto de F\'{\i}sica, Universidad Nacional Aut\'{o}noma de M\'{e}xico, Mexico City, Mexico
\\
$^{65}$Institut Pluridisciplinaire Hubert Curien (IPHC), Universit\'{e} de Strasbourg, CNRS-IN2P3, Strasbourg, France
\\
$^{66}$iThemba LABS, National Research Foundation, Somerset West, South Africa
\\
$^{67}$Joint Institute for Nuclear Research (JINR), Dubna, Russia
\\
$^{68}$Konkuk University, Seoul, South Korea
\\
$^{69}$Korea Institute of Science and Technology Information, Daejeon, South Korea
\\
$^{70}$KTO Karatay University, Konya, Turkey
\\
$^{71}$Laboratoire de Physique Corpusculaire (LPC), Clermont Universit\'{e}, Universit\'{e} Blaise Pascal, CNRS--IN2P3, Clermont-Ferrand, France
\\
$^{72}$Laboratoire de Physique Subatomique et de Cosmologie, Universit\'{e} Grenoble-Alpes, CNRS-IN2P3, Grenoble, France
\\
$^{73}$Laboratori Nazionali di Frascati, INFN, Frascati, Italy
\\
$^{74}$Laboratori Nazionali di Legnaro, INFN, Legnaro, Italy
\\
$^{75}$Lawrence Berkeley National Laboratory, Berkeley, California, United States
\\
$^{76}$Moscow Engineering Physics Institute, Moscow, Russia
\\
$^{77}$Nagasaki Institute of Applied Science, Nagasaki, Japan
\\
$^{78}$National Centre for Nuclear Studies, Warsaw, Poland
\\
$^{79}$National Institute for Physics and Nuclear Engineering, Bucharest, Romania
\\
$^{80}$National Institute of Science Education and Research, Bhubaneswar, India
\\
$^{81}$National Research Centre Kurchatov Institute, Moscow, Russia
\\
$^{82}$Niels Bohr Institute, University of Copenhagen, Copenhagen, Denmark
\\
$^{83}$Nikhef, Nationaal instituut voor subatomaire fysica, Amsterdam, Netherlands
\\
$^{84}$Nuclear Physics Group, STFC Daresbury Laboratory, Daresbury, United Kingdom
\\
$^{85}$Nuclear Physics Institute, Academy of Sciences of the Czech Republic, \v{R}e\v{z} u Prahy, Czech Republic
\\
$^{86}$Oak Ridge National Laboratory, Oak Ridge, Tennessee, United States
\\
$^{87}$Petersburg Nuclear Physics Institute, Gatchina, Russia
\\
$^{88}$Physics Department, Creighton University, Omaha, Nebraska, United States
\\
$^{89}$Physics Department, Panjab University, Chandigarh, India
\\
$^{90}$Physics Department, University of Athens, Athens, Greece
\\
$^{91}$Physics Department, University of Cape Town, Cape Town, South Africa
\\
$^{92}$Physics Department, University of Jammu, Jammu, India
\\
$^{93}$Physics Department, University of Rajasthan, Jaipur, India
\\
$^{94}$Physikalisches Institut, Eberhard Karls Universit\"{a}t T\"{u}bingen, T\"{u}bingen, Germany
\\
$^{95}$Physikalisches Institut, Ruprecht-Karls-Universit\"{a}t Heidelberg, Heidelberg, Germany
\\
$^{96}$Physik Department, Technische Universit\"{a}t M\"{u}nchen, Munich, Germany
\\
$^{97}$Purdue University, West Lafayette, Indiana, United States
\\
$^{98}$Pusan National University, Pusan, South Korea
\\
$^{99}$Research Division and ExtreMe Matter Institute EMMI, GSI Helmholtzzentrum f\"ur Schwerionenforschung, Darmstadt, Germany
\\
$^{100}$Rudjer Bo\v{s}kovi\'{c} Institute, Zagreb, Croatia
\\
$^{101}$Russian Federal Nuclear Center (VNIIEF), Sarov, Russia
\\
$^{102}$Saha Institute of Nuclear Physics, Kolkata, India
\\
$^{103}$School of Physics and Astronomy, University of Birmingham, Birmingham, United Kingdom
\\
$^{104}$Secci\'{o}n F\'{\i}sica, Departamento de Ciencias, Pontificia Universidad Cat\'{o}lica del Per\'{u}, Lima, Peru
\\
$^{105}$Sezione INFN, Bari, Italy
\\
$^{106}$Sezione INFN, Bologna, Italy
\\
$^{107}$Sezione INFN, Cagliari, Italy
\\
$^{108}$Sezione INFN, Catania, Italy
\\
$^{109}$Sezione INFN, Padova, Italy
\\
$^{110}$Sezione INFN, Rome, Italy
\\
$^{111}$Sezione INFN, Trieste, Italy
\\
$^{112}$Sezione INFN, Turin, Italy
\\
$^{113}$SSC IHEP of NRC Kurchatov institute, Protvino, Russia
\\
$^{114}$Stefan Meyer Institut f\"{u}r Subatomare Physik (SMI), Vienna, Austria
\\
$^{115}$SUBATECH, Ecole des Mines de Nantes, Universit\'{e} de Nantes, CNRS-IN2P3, Nantes, France
\\
$^{116}$Suranaree University of Technology, Nakhon Ratchasima, Thailand
\\
$^{117}$Technical University of Ko\v{s}ice, Ko\v{s}ice, Slovakia
\\
$^{118}$Technical University of Split FESB, Split, Croatia
\\
$^{119}$The Henryk Niewodniczanski Institute of Nuclear Physics, Polish Academy of Sciences, Cracow, Poland
\\
$^{120}$The University of Texas at Austin, Physics Department, Austin, Texas, United States
\\
$^{121}$Universidad Aut\'{o}noma de Sinaloa, Culiac\'{a}n, Mexico
\\
$^{122}$Universidade de S\~{a}o Paulo (USP), S\~{a}o Paulo, Brazil
\\
$^{123}$Universidade Estadual de Campinas (UNICAMP), Campinas, Brazil
\\
$^{124}$Universidade Federal do ABC, Santo Andre, Brazil
\\
$^{125}$University of Houston, Houston, Texas, United States
\\
$^{126}$University of Jyv\"{a}skyl\"{a}, Jyv\"{a}skyl\"{a}, Finland
\\
$^{127}$University of Liverpool, Liverpool, United Kingdom
\\
$^{128}$University of Tennessee, Knoxville, Tennessee, United States
\\
$^{129}$University of the Witwatersrand, Johannesburg, South Africa
\\
$^{130}$University of Tokyo, Tokyo, Japan
\\
$^{131}$University of Tsukuba, Tsukuba, Japan
\\
$^{132}$University of Zagreb, Zagreb, Croatia
\\
$^{133}$Universit\'{e} de Lyon, Universit\'{e} Lyon 1, CNRS/IN2P3, IPN-Lyon, Villeurbanne, Lyon, France
\\
$^{134}$Universit\`{a} di Brescia, Brescia, Italy
\\
$^{135}$V.~Fock Institute for Physics, St. Petersburg State University, St. Petersburg, Russia
\\
$^{136}$Variable Energy Cyclotron Centre, Kolkata, India
\\
$^{137}$Warsaw University of Technology, Warsaw, Poland
\\
$^{138}$Wayne State University, Detroit, Michigan, United States
\\
$^{139}$Wigner Research Centre for Physics, Hungarian Academy of Sciences, Budapest, Hungary
\\
$^{140}$Yale University, New Haven, Connecticut, United States
\\
$^{141}$Yonsei University, Seoul, South Korea
\\
$^{142}$Zentrum f\"{u}r Technologietransfer und Telekommunikation (ZTT), Fachhochschule Worms, Worms, Germany
\endgroup